# Searching for the neurite density with diffusion MRI: challenges for biophysical modeling

Björn Lampinen[a], Filip Szczepankiewicz[a,b], Mikael Novén[c], Danielle van Westen[d], Oskar Hansson[e], Elisabet Englund[f], Johan Mårtensson[g], Carl-Fredrik Westin[b] and Markus Nilsson[d]

[a]Clinical Sciences Lund, Medical Radiation Physics, Lund University, Lund, Sweden; [b]Brigham and Women's Hospital, Harvard Medical School, Boston, MA, United States; [c]Centre for Languages and Literature, Lund University, Lund, Sweden; [d]Clinical Sciences Lund, Radiology, Lund University, Lund, Sweden; [e]Clinical Sciences Malmö, Clinical Memory Research Unit, Lund University, Lund, Sweden; [f]Clinical Sciences Lund, Oncology and Pathology, Lund University, Lund, Sweden; [g]Clinical Sciences, Department of Logopedics, Phoniatrics, and Audiology, Lund University, Lund, Sweden

**Abstract** – In vivo mapping of the neurite density with diffusion MRI (dMRI) is a high but challenging aim. First, it is unknown whether all neurites exhibit completely anisotropic ('stick-like') diffusion. Second, the 'density' of tissue components may be confounded by non-diffusion properties such as T2 relaxation. Third, the domain of validity for the estimated parameters to serve as indices of neurite density is incompletely explored. We investigated these challenges by acquiring data with 'b-tensor encoding' and multiple echo times in both healthy brain and white matter lesions. Results showed that microscopic anisotropy from b-tensor data is associated with myelinated axons but not with dendrites. Furthermore, b-tensor and multi-echo data showed that unbiased density estimates in white matter lesions require data-driven estimates of compartment-specific T2 times. Finally, the 'stick' fractions of different biophysical models could generally not serve as neurite density indices across the healthy brain and white matter lesions, where outcomes of comparisons depended on the choice of constraints. In particular, constraining compartment-specific T2 times was ambiguous in the healthy brain and had a large impact on estimated values. In summary, estimating neurite density may require accounting for different diffusion and/or T2 properties between axons and dendrites. Constrained 'index' parameters could be valid within limited domains that should be delineated by future studies.

## 1. Introduction

Diffusion MRI (dMRI) makes use of the micrometer scale displacement of water molecules to detect microstructural alterations in the brain due to disease (Moseley et al. 1990, Chenevert et al. 2000, Werring et al. 2000, Horsfield et al. 2002, Van Cauter et al. 2012, Surova et al. 2013) as well as normal processes such as learning (Scholz et al. 2009, Sagi et al. 2012) and maturation (Lebel et al. 2008, Löbel et al. 2009). However, dMRI does not yield tissue quantities directly. For example, the fractional anisotropy (FA) parameter of diffusion tensor imaging (DTI) (Basser et al. 1994) cannot be directly associated to the density of anisotropic structures such as axons since it is sensitive to other factors such as their orientation dispersion (Pierpaoli et al. 1996, Douaud et al. 2011).

Biophysical models seek to increase the specificity of dMRI by dividing the signal between 'compartments' to estimate tissue quantities (Nilsson et al. 2013, Novikov et al. 2016, Alexander et al. 2017, Kiselev 2017, Novikov et al. 2018a). The target quantity of early approaches was the axonal volume fraction in coherent white matter, characterized by diffusion in highly anisotropic structures such as cylinders (Assaf et al. 2005), with a small but non-zero (apparent) radial diffusivity, or 'sticks' (Behrens et al. 2003) with an (apparent) radial diffusivity of zero. Further developments extended models to non-coherent white matter by incorporating an orientation distribution function (Jespersen et al. 2007, Kaden et al. 2007, Zhang et al. 2011, Sotiropoulos et al. 2012, Zhang et al. 2012, Tariq et al. 2016) or by mimicking complete orientation dispersion (Kroenke et al. 2004) through so-called powder-averaging. Several contemporary approaches aspire to encompass also gray matter and to capture the combined volume fractions of axons and dendrites (neurites) in a single 'neurite density' parameter (the 'stick' fraction) (Jespersen et al. 2007, Zhang et al. 2012, White et al. 2013, Kaden et al. 2016). However, estimating a specific tissue quantity such as neurite density from clinical dMRI data necessitates simplifying assumptions whose accuracy and generality are unknown. This work focuses on three challenges for scientists using biophysical models to estimate the neurite density.

The first challenge is that mapping neurites requires a correct model for diffusion in not only axons but also in dendrites. Biophysical models typically use the 'neurite assumption' of representing both these structures in a single 'stick' compartment with fully anisotropic (one dimensional) diffusion. The rationale is that axons and dendrites should exhibit a radial diffusivity of approximately zero (Behrens et al. 2003, Kroenke et al. 2004) due to their approximately cylindrical shape with diameters that are smaller than the resolution limit (Nilsson et al. 2017). Studies have confirmed this property for axons (McKinnon et al. 2017, Veraart et al. 2017a) but the evidence is less conclusive for dendrites. Although 'stick-like' diffusion has been demonstrated in cortical gray matter (Novikov et al. 2014), this could merely reflect the presence of myelinated axons, as suggested by the study on fixed tissue using histology by (Jespersen et al. 2010). Furthermore, the low levels of microscopic anisotropy observed in gray matter compared to white matter (Lampinen et al. 2017, Lawrenz et al. 2018, Novikov et al. 2018b) may indicate that the radial diffusivity of dendrites is actually non-zero.

The second challenge is that mapping densities (volume fractions) from signal fractions requires accounting for potential T2 relaxation time differences between tissue components. Since T2 times cannot be estimated from dMRI data acquired with a single echo time (TE), contemporary approaches implicitly use the 'density assumption' that these are equal between 'neurite' and 'non-neurite' tissue. Some evidence suggests that such T2 differences are indeed small in white matter, where intra- and extra-axonal water pools are not clearly distinguishable based on multi-exponential T2 relaxation (Mackay et al. 1994, Whittall et al. 1997) and where studies have reported negligible effects on diffusion from varying TE (Beaulieu et al. 1998, Clark et al. 2000). The issue is not well investigated in gray matter, however, and more recent white matter studies indicate that the T2 time of intra-axonal water may be longer than that of extra-axonal water (De Santis et al. 2016, Veraart et al. 2017b). Furthermore, the intra- and extra-neurite T2 properties should be perturbed in pathologies involving an enlargement of the extracellular space such as ischemic degeneration (Englund et al. 1990) or the vascular edemas associated with blood-brain barrier damage in conditions such as tumors and acute MS lesions (Ballabh et al. 2004).

The third challenge is to find the range of conditions, or 'domain of validity', where the specific interpretations of biophysical model parameters are accurate. This domain may be both small and difficult to delineate for a parameter that is interpreted as the actual physical neurite density (fractional voxel volume occupied by neurites). The domain may be widened, however, if the parameter as interpreted as an 'index' for neurite density, as in (Zhang et al. 2012, Kaden et al. 2016), since this acknowledges a multifactorial origin and only claims the 'ordinal accuracy' to rank observations. Even for neurite density indices, however, the domain of validity must be delineated to avoid conditions where factors such as T2 relaxation are sufficiently large to 'break' the index property and confound the outcome of comparisons.

This work investigates the three above challenges through a 'multidimensional' diffusion acquisition using the novel 'b-tensor encoding' approach (Eriksson et al. 2013, Lasič et al. 2014, Eriksson et al. 2015) as well as multiple echo times (de Almeida Martins et al.

2018). First, we examined whether mapping neurites based on the property of microscopic diffusion anisotropy is supported by data. Using the high specificity of b-tensor encoding and knowledge from histology literature, we compared the anisotropy between multiple healthy brain regions that are substantially different with respect to axonal density but similar with respect to the combined axonal and dendritic (neurite) density. This tests a prediction of the neurite assumption, that microscopic anisotropy should reflect neurite density, against that of an alternative hypothesis, that this property is mainly due to axons. Second, we explored whether T2 relaxation is likely to confound density estimates in healthy white and gray matter as well as in white matter lesions. Using the combined b-tensor and multi-echo data, we attempted independent estimation of the 'stick' fraction and compartment-specific T2 times in a minimally constrained biophysical model. This tests the feasibility of disentangling density from T2 relaxation and addresses the validity of the density assumption. White matter lesions are suitable in this context since they may exhibit changes to both diffusion anisotropy, through demyelination (Swieten et al. 1991) and axonal loss (Englund and Brun 1990) and to T2 relaxation (Englund et al. 1987). Third, we tested whether the domain of validity for neurite density indices encompass healthy white and gray matter as well as white matter lesions. Using models that are prevalent in the literature, we compared the resulting rankings of a range of brain regions with respect to the 'stick' fraction. Since there can be only one true ranking with respect to a quantity such as neurite density, disagreement between models signifies a domain of validity violation. The results have impact on the standard interpretation of dMRI data, which is elaborated on in the discussion.

## 2. Theory

The techniques for data analysis described in this section were employed after arithmetic averaging of the signal across diffusion-encoding directions, so-called 'powder averaging' (Callaghan et al. 1979, Jespersen et al. 2013, Lasič et al. 2014). Provided data is acquired with a sufficient number of directions (Szczepankiewicz et al. 2016c), powder averaging yields a signal whose orientation-invariant aspects of diffusion are preserved but with an orientation distribution that mimics complete dispersion of anisotropic structures.

### 2.1 Representation-based analysis

We represent the powder-averaged b-tensor data with a cumulant expansion in b-values, according to

$$\log(S) \approx \log(S_0) - b \cdot \mathrm{MD} + b^2 \cdot (\mathrm{MK_I} + b_\Delta^2 \cdot \mathrm{MK_A}) \cdot \mathrm{MD}^2 / 6. \qquad \text{Eq. 1}$$

Four parameters represent the signal: the non-diffusion-weighted signal ($S_0$), the mean diffusivity (MD), the 'isotropic kurtosis' ($\mathrm{MK_I}$) and the 'anisotropic kurtosis' ($\mathrm{MK_A}$). Two parameters describe the experiment: the 'size' ($b$) and 'shape' ($b_\Delta$) parameters of the b-tensor, where the former is the conventional b-value (Le Bihan et al. 1986) and the latter ranges from −0.5 (planar) through zero (spherical) to unity (linear) (Eriksson et al. 2015, Topgaard 2016, Topgaard 2017). Under the assumption that the diffusion process can be described by a set of non-exchanging compartments with approximately Gaussian diffusion, the $\mathrm{MK_I}$ parameter represents variance in isotropic diffusivities ('isotropic heterogeneity') and the $\mathrm{MK_A}$ parameter represents microscopic diffusion anisotropy (Szczepankiewicz et al. 2016b, Westin et al. 2016, Topgaard 2017). These properties cannot be separated with data acquired with a single shape of the b-tensor (Mitra 1995, Topgaard 2017). For example, using only the linear tensor encoding (LTE, $b_\Delta = 1$) of conventional multi-shell dMRI yields the (powder-averaged) signal equation of diffusion kurtosis imaging (DKI) (Yablonskiy et al. 2003, Jensen et al. 2005) and three parameters: $S_0$, MD and MK, where MK = $\mathrm{MK_I}$ + $\mathrm{MK_A}$, as a special case. On the other hand, data acquired with two or more b-tensor shapes, such as the combination of LTE and spherical tensor encoding (STE; $b_\Delta = 0$) (Lasič et al. 2014, Szczepankiewicz et al. 2015, Szczepankiewicz et al. 2016b) or LTE and planar tensor encoding (PTE; $b_\Delta = -\frac{1}{2}$) (Jespersen et al. 2013, Lawrenz et al. 2013), allows estimation of four parameters and disambiguating the source for kurtosis.

*2.2 Model-based analysis*

Biophysical models express the signal as the sum of signals from different compartments, according to

$$S = \Sigma S_k = \Sigma f_k \cdot S_{PD;k} \cdot A_{T1;k} \cdot A_{T2;k} \cdot A_{D;k}, \qquad \text{Eq. 2}$$

where, for the k:th compartment, $f_k$ is the fraction ($\Sigma f_k = 1$), $S_{PD;k}$ is the proton density-weighted signal, $A_{T1;k}$ and $A_{T2;k}$ are the attenuations due to T1 and T2 relaxation, respectively, and $A_{D;k}$ is the attenuation due to diffusion. For the powder-averaged signal, the diffusion attenuation is solely a function of orientation-invariant aspects of the diffusion and the encoding. We represent compartment diffusion with an axisymmetric diffusion tensor described by its 'size' (isotropic diffusivity), $D_I = 1/3 \cdot D_\parallel + 2/3 \cdot D_\perp$ and 'shape' (anisotropy), $D_\Delta = (D_\parallel - D_\perp) / (D_\parallel + 2 \cdot D_\perp) \in [-\frac{1}{2}\ 1]$, where $D_\parallel$ and $D_\perp$ are the axial and radial diffusivities, respectively As we similarly represent diffusion encoding by an axisymmetric b-tensor, parameterized by $b$ and $b_\Delta$, the compartment diffusion attenuation is a function of four scalar parameters, according to (Eriksson et al. 2015, Topgaard 2016, Topgaard 2017)

$$A_{D;k}(b, b_\Delta, D_{I;k}, D_{\Delta;k}) = \exp(-bD_{I;k}\ [1 - b_\Delta D_{\Delta;k}]) \cdot g(3bD_{I;k}b_\Delta D_{\Delta;k}), \qquad \text{Eq. 3}$$

where

$$g(\alpha) = \int_0^1 \exp(-\alpha x^2)\mathrm{d}x = \sqrt{\frac{\pi}{4\alpha}}\mathrm{erf}\ (\sqrt{\alpha}), \qquad \text{Eq. 4}$$

and erf($x$) is the error function (Callaghan et al. 1979).

We define a minimally constrained biophysical model based on Eq. 2 using three compartments: a 'ball' (B), a 'stick' (S) and cerebrospinal fluid (CSF; C). By design, the 'ball' and CSF compartments feature isotropic diffusion ($D_{\Delta;B} = 0$, $D_{\Delta;C} = 0$) while the 'stick' compartment features completely anisotropic diffusion ($D_{\Delta;S} = 1$). While proton densities and T1 relaxation times are assumed to be equal for all compartments, the T2 times are free for 'balls' (T2$_B$) and 'sticks' (T2$_S$). The T2 time of CSF was fixed as T2$_C$ = 1400 ms, motivated by results from Weigel et al. (2006) and Hopkins et al. (1986). Finally, the isotropic diffusivities of 'balls' ($D_{I;B}$) and 'sticks' ($D_{I;S}$) are free parameters, while the isotropic diffusivity of CSF is fixed as $D_{I;C}$ = 3 μm$^2$/ms. In total, the model features seven free parameters, summarized in Table 1: $S_0$, $f_S$ (the 'stick' fraction), $f_C$ (the CSF fraction), T2$_B$, T2$_S$, $D_{I;B}$ and $D_{I;S}$. Together with the experimental parameters, $b$, $b_\Delta$ and TE these combine to form the full signal equation, according to

$$S = S_0 \cdot [(1 - f_S - f_C) \cdot A_{T2;B} \cdot A_{D;B} + f_S \cdot A_{T2;S} \cdot A_{D;S} + f_C \cdot A_{T2;C} \cdot A_{D;C})], \qquad \text{Eq. 5}$$

where the attenuations are given by

$$A_{T2;B/S/C} = \exp(-TE/T2_{B/S/C}), \qquad\qquad\qquad\qquad \text{Eq. 6}$$

$$A_{D;B/C} = \exp(-bD_{I;B/C}), \qquad\qquad\qquad\qquad\qquad \text{Eq. 7}$$

$$A_{D;S} = \exp(-bD_{I;S}[1 - b_\Delta]) \cdot g(3bD_{I;S}b_\Delta), \qquad\quad \text{Eq. 8}$$

using $g$ from Eq 4. This model thus resembles the 'standard model' for white matter (Novikov et al. 2016, Novikov et al. 2018b) but extended to gray matter (by the CSF fraction) and to feature T2 relaxation. Unlike the standard model, however, we employ the 'ball' constraint of isotropic diffusion outside of 'stick-like' structures. The choice was made to improve feasibility of model fitting, but limited the subsequent data analysis to regions with a high orientation dispersion where we assume the extra-axonal diffusion to be nearly isotropic.

We define six sets of constraints on top of the minimally constrained model (set C0; Table 1), yielding six nested 'submodels' that reflect contemporary approaches to obtain 'stick' fractions. Set C1 yields an approximation of the spherical mean technique (SMT) model (Kaden et al. 2016), but extended to feature two free compartment T2 times. It comprises a CSF constraint, given by

$$f_C = 0, \qquad\qquad\qquad\qquad\qquad\qquad\qquad\qquad\qquad \text{Eq. 9}$$

and a constraint that calculates the 'ball' isotropic diffusivity from the 'stick' isotropic diffusivity and the 'stick' fraction, according to

$$D_{I;B} = D_{I;S} \cdot (3 - 2f_S). \qquad\qquad\qquad\qquad\qquad \text{Eq. 10}$$

Set C2 yields a powder-averaged equivalent of the Ball and stick (Behrens et al. 2003) and Ball and rackets (Sotiropoulos et al. 2012) models, but extended to feature free compartment diffusivities. It comprises a T2 constraint, given by

$$T2_B = T2_S, \qquad\qquad\qquad\qquad\qquad\qquad\qquad\quad \text{Eq. 11}$$

and the CSF constraint (Eq. 9). Set C3 yields an approximation of the SMT model and comprises all the above constraints (Eqs. 9–11). Set C4 yields the CODIVIDE model (Lampinen et al. 2017) and comprises the T2 constraint (Eq. 11) and a diffusivity constraint, given by

$$D_{I;B} = D_{I;S}. \qquad \text{Eq. 12}$$

Set C5 approximates the powder-averaged NODDI model (Zhang et al. 2012, Lampinen et al. 2017) and comprises the relation between isotropic diffusivities and the 'stick' fraction in Eq. 10, the T2 constraint (Eq. 11), and a fixed 'stick' isotropic diffusivity, given by

$$D_{I;S} = 0.57 \ \mu m^2/ms. \qquad \text{Eq. 13}$$

Set C6 equals set C2 but uses a T2 constraint different than Eq. 11, given by

$$T2_S = 70 \ ms, \qquad \text{Eq. 14}$$

a choice inspired by previous estimates of the voxel T2 time in white matter at 3T (Whittall et al. 1997, Wansapura et al. 1999) and results in white matter lesions obtained in this work. Note that compared to SMT and NODDI, sets C3 and C5 use the 'ball' constraint ($D_\Delta = 0$) for the 'non-stick' compartment rather than defining its shape from a tortuosity relation (Szafer et al. 1995). The difference was negligible in the herein investigated regions (supporting information). Finally, the original NODDI model employs a slight rearrangement of the 'stick' and CSF fractions (Zhang et al. 2012).

While a compartment-based model (Eq. 2) can attain any capacity, using the word as defined in Goodfellow et al. (2016), it is generally necessary to constrain both the number of compartments and the compartment properties to match the degrees of freedom of data and avoid degeneracy in parameter estimation (Jelescu et al. 2016a).

One type of constraint concerns compartment relaxation times (and proton densities), which cannot be estimated unless the dMRI acquisition features additional encoding dimensions with sensitivity to these properties. Most biophysical models are not designed for variable relaxation weighting and typically use the density assumption, given by

$$S_{PD;k} \cdot A_{T1;k} \cdot A_{T2;k} = S_0, \text{ for all } k, \qquad \text{Eq. 15}$$

as in sets C2–C5 (with a different T2 for CSF). The dMRI data obtained for multiple TE used here, however, allow tentatively releasing the constraint of equal 'ball' and 'stick' T2 times (Eq. 11) in the minimally constrained model. Our multi-echo data were acquired for $b \leq 0.5$ ms/µm², thus probing the TE dependence of the first cumulant of the diffusion attenuation (the mean diffusivity, MD) and yielding the following relation

$$\text{MD(TE)} = [1 - f_S\text{'(TE, T2}_B, \text{T2}_S)] \cdot D_{I;B} + f_S\text{'(TE, T2}_B, \text{T2}_S) \cdot D_{I;S}, \qquad \text{Eq. 16}$$

where $f_S\text{'} = [S_S/S]_{b=0}$ is the relaxation-weighted 'stick' signal fraction before diffusion attenuation. For simplicity, Eq. 16 assumes $f_C = 0$. Together with the voxel T2 time, this relation may allow estimation of both T2$_B$ and T2$_S$. However, the gained precision depends on the difference between 'ball' and 'stick' isotropic diffusivities, which is clear when considering that the change in MD with respect to TE,

$$\text{dMD/dTE} = \text{d}f_S\text{'/dTE} \cdot (D_{I;S} - D_{I;B}), \qquad \text{Eq. 17}$$

approaches zero as $D_{I;S}$ approaches $D_{I;B}$.

Another type of constraint concerns diffusion properties. Conventional multi-shell dMRI only allows the estimation of two diffusion parameters (Kiselev et al. 2007), extended to three if multiple shapes of the b-tensor are used (Lasič et al. 2014, Lampinen et al. 2017). Compared to the minimally constrained model (set C0), which features four diffusion parameters ($f_S, f_C, D_{I;B}$ and $D_{I;S}$), the constraint sets designed for b-tensor encoding (C2, C4 and C6) use one diffusion constraint and leave three parameters free. Conversely, constraint sets designed for conventional multi-shell dMRI (C2, C4 and C6) use two diffusion constraints and leave two parameters free. Finally, set C1 was designed for b-tensor encoding but still uses two diffusion constraints. Instead of estimating a third diffusion parameter, set C1 relates diffusion to relaxation through the relations in Eq. 10 and Eq. 16.

**Table 1** – Parameters and sets of constraints for the model-based analysis.

| Constraint set | $n_{par}$ | $S_0$ | $f_S$ | $f_C$ | T2$_B$ | T2$_S$ | $D_{I;B}$ | $D_{I;S}$ |
|---|---|---|---|---|---|---|---|---|
| C0 (minimally constrained) | 7 | - | - | - | - | - | - | - |
| C0 iterated non-cortex | 5 | - | iterated[a] | 0 | - | - | - | - |
| C0 iterated cortex | 6 | - | iterated[a] | - | - | - | - | - |
| C1 (SMT[b,c]) | 5 | - | - | 0 | - | - | Eq. 10 | - |
| C2 (Ball and stick/rackets[d,e]) | 5 | - | - | 0 | - | T2$_B$ | - | - |
| C3 (SMT[c]) | 4 | - | - | 0 | - | T2$_B$ | Eq. 10 | - |
| C4 (CODIVIDE) | 5 | - | - | - | - | T2$_B$ | - | $D_{I;B}$ |
| C5 (NODDI[c,e]) | 4 | - | - | - | - | T2$_B$ | Eq. 10 | 0.57 |
| C6 (Ball and stick/rackets[d,e,f]) | 5 | - | - | 0 | - | 70 | - | - |

Relaxation times are in ms and diffusivities are in $\mu m^2/ms$; the '-' denote free parameters; [a]$f_S \in$ [0, 0.0125, 0.0375, ..., 0.9875]; [b]featuring two free T2 parameters; [c]with a 'ball-shaped' ($D_\Lambda = 0$) 'non-stick' compartment; [d]featuring two free isotropic diffusivity parameters; [e]powder-averaged; [f]using a different T2 constraint.

## 3. Methods

### 3.1 Subjects, acquisition and post-processing

Data were acquired on a Siemens MAGNETOM Prisma 3T system (Siemens Healthcare, Erlangen, Germany) in three subject groups: A ($n = 5$; age $= 32 \pm 4$ years; all male), B ($n = 20$; age $= 25 \pm 4$ years; male/female $= 11/9$), and C ($n = 10$; age $= 70 \pm 10$ years; male/female $= 5/5$). The group C data were included to study white matter lesions and came from a cohort of Parkinson patients and controls. The selection criteria were related to white matter lesion status and disregarded patient/control status.

We performed dMRI with b-tensor encoding using a prototype echo-planar diffusion sequence (Szczepankiewicz et al. 2016a) with encoding gradient waveforms optimized for minimal TE (Sjölund et al. 2015). Data were acquired using both linear and spherical b-tensors yielding linear tensor encoding (LTE) and spherical tensor encoding (STE), respectively. Groups A and C were additionally scanned with a low b-value multi-echo sequence using LTE. T1-weighted MPRAGE images were acquired for segmentation purposes. The experiments were approved by the regional ethical review board in Lund, and all subjects gave informed consent. Sequence parameters for all diffusion acquisitions are shown in Table 2.

The dMRI data were corrected for eddy-currents and subject motion using ElastiX (Klein et al. 2010) with extrapolated target volumes (Nilsson et al. 2015). Before model fitting, data were arithmetically averaged across diffusion-encoding directions (powder-averaged) for each b-value and each b-tensor shape. For the purpose of ROI definition and image registration, we performed a DTI analysis of LTE data at $b \leq 1$ ms/μm². All processing was performed using software available in the multidimensional dMRI toolbox (https://github.com/markus-nilsson/md-dmri) (Nilsson et al. 2018).

### 3.2 Regions of interest (ROIs)

The study used data from 13 bilateral regions of interest (ROIs; Fig. 1), representing white matter, with or without lesions, together with deep and cortical gray matter. The gray matter ROIs were selected to maximize the expected difference in axonal content. Since histology literature provides little information on axonal content variation in gray matter, regions were selected based on myelin content, which is presumed to be closely related.

For group A, five ROIs were defined to represent white matter by the anterior corona radiata, deep gray matter by the high myelin ventrolateral thalamus and the low myelin mediodorsal thalamus (Danos et al. 2003), and cortical gray matter by the high myelin primary auditory cortex (Hopf 1955, Bock et al. 2009) and the low myelin insular cortex (Öngür et al. 2003). For group B, eight cortical ROIs were defined to represent high myelin cortex by the primary auditory cortex, the primary visual cortex (Bock et al. 2009), the primary motor cortex (Hopf 1956) and the posterior cingulate cortex (Morris et al. 2000), and low myelin cortex by the insular cortex, the medial orbitofrontal cortex (Hopf 1956), the polar plane (Hopf 1955), and the anterior cingulate cortex (Öngür et al. 2003). For group C, the anterior corona radiata was defined as in group A. For each of the ten subjects, this (bilateral) region was classified as either 'normal-appearing' (NA) or 'lesion', resulting in five subjects per category.

ROI definition was performed manually for groups A and C, using the $S_0$ and directionally encoded color (DEC) maps from DTI, and automatically for group B, using the Desikan-Killany and Destrieux parcellations of Freesurfer. The automatic procedure involved registering the $S_0$ image from DTI to the MPRAGE using rigid-body registration implemented in ElastiX. A visual inspection assessed that the registrations yielded proper alignments and that the resulting ROIs were located within the cortex and minimally affected by EPI distortions. DTI parameters were used to make final adjustments to these ROIs. To reduce contamination with cerebrospinal fluid (CSF) and white matter, we excluded voxels with a high mean diffusivity (MD > 1.1 $\mu m^2$/ms) and a high orientation coherence (FA > 0.5). The signal was averaged across voxels when extracted, for each ROI, before further analysis.

*3.3 Brain patterns of microscopic anisotropy from b-tensor encoding*

The level of microscopic anisotropy was assessed for all subjects and ROIs (Fig. 1) by the $MK_A$ parameter obtained from the combined LTE and STE data (Table 2) in the representation-based analysis using Eq. 1. Paired two-tailed *t*-tests were performed comparing $MK_A$ values between the high myelin ventrolateral thalamus and the low myelin mediodorsal thalamus ROIs in group A, and between the low and high myelin cortical regions ROIs in group B. In the latter comparison, the signal was pooled, for each subject, across the four ROIs representing either high or low myelin. In addition, we compared the contrast of the $MK_A$ maps with a myelin-stained coronal brain section obtained from a subject outside of the imaged group. A 6 μm thick section was cut at the level of the anterior thalamus and stained for myelin using Luxol fast blue together with Cresyl violet counter staining of nuclear structures.

*3.4 Biophysical modeling with compartment-specific T2 times*

The minimally constrained model (Eq. 5) was applied to the combined b-tensor and multi-echo data from the group A and C ROIs (Fig. 1). We first assessed the model's precision to estimate the 'stick' fraction together with compartment-specific T2 times, and then extracted parameter values from regions where precision was deemed sufficient.

Model precision was assessed by plotting a metric of goodness of fit against different values of the 'stick' fraction, fixed prior to the fitting in 40 steps between zero and one. If data determine all model parameters, the plot should exhibit a clear minimum for some 'stick' fraction. If data are insufficient, however, multiple 'stick' fractions should yield a similarly good fit and result in a flatter plot. The CSF fraction was fixed to zero in the non-cortical regions. The CSF fraction was free in the cortex, reducing the precision, but the procedure still yielded a good representation of the range of data-compatible 'stick' fractions (supporting information).

Goodness of fit was defined as the normalized residual variance (NRV) obtained by dividing the residual variance ($\sigma_R^2$) by the signal variance due to noise ($\sigma^2_{noise}$), according to

$$NRV = \sigma_R^2/\sigma^2_{noise} = [\Sigma n_{pa;i} (S_i - S'_i)^2 / (n - k)] / (\sigma^2_{noise} / n_{voxel}) \qquad Eq. 18$$

where $n_{pa;i}$ is the number of powder-averaged directions for the i:th combination of measurement parameters ($b$, $b_\Delta$, and TE), $S_i$ and $S'_i$ are the powder-averaged signals obtained from the measurement and predicted by the fit, respectively, $n$ is the total number of samples (Table 2) and $k$ is the number of parameters that were free to vary in the fitting (Table 1). The value of $\sigma^2_{noise}$ was estimated for each ROI as the voxel and b-value average of the STE signal variance across the multiple repetitions.

For group C, the model was extended to feature the tissue T1 time ($T1_{B/S} = T1$) and the constraint $T1_C = 5700$ ms to account for the multiple repetition times of the combined b-tensor and multi-echo data (Table 2). The precision in the estimated tissue T1 time was likely low given the rather long repetition times used in both sequences, but it was included as a nuisance parameter.

*3.5 Brain region rankings by the 'stick' fractions of constrained biophysical models*

The six constrained 'submodels' of the minimally constrained model (Table 1) were fitted to the combined b-tensor and multi-echo data in the group A and C ROIs (Fig. 1). First, we ensured that the constrained models yielded precise solutions by plotting the goodness of fit (Eq. 18) against fixed values of the 'stick' fraction. Second, we compared how each model ranked the different regions with respect to the 'stick' fraction, treating the anterior corona radiata with or without lesion from group C as separate regions. Third, we assessed whether the sets of constraints were compatible with data by comparing their corresponding model's fit quality in the different regions, defined as the sum of squared residuals (SSR) pooled from the combined fits in all subjects.

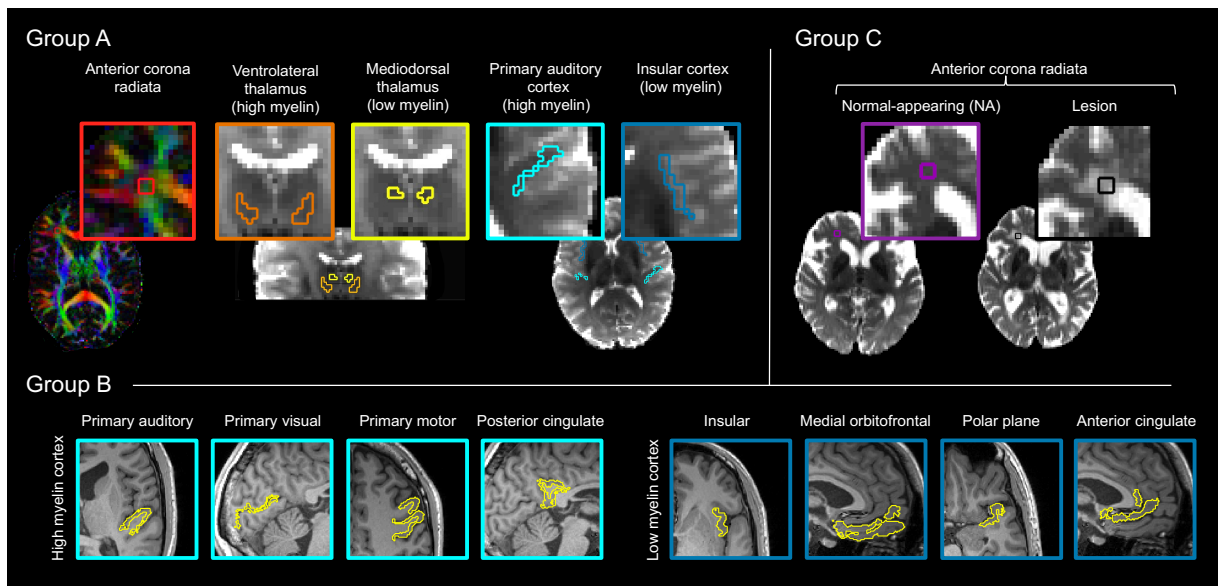

**Figure 1 – Regions of interest (ROIs).** The study used 13 bilateral ROIs, representing white matter with or without lesions, together with gray matter featuring either high or low myelin content based on histology literature. In Group A ($n = 5$), we manually defined five ROIs: white matter (anterior corona radiata), deep gray matter with high myelin (ventrolateral thalamus) and low myelin (mediodorsal thalamus), and cortical gray matter with high myelin (primary auditory cortex) and low myelin (insular cortex). In Group B ($n = 20$), we used the Desikan-Killany and Destrieux parcellations of Freesurfer to define eight cortical ROIs: four with high myelin (primary auditory cortex, primary visual cortex, primary motor cortex and posterior cingulate cortex) and four with low myelin (insular cortex, medial orbitofrontal cortex, the polar plane and anterior cingulate cortex). In group C ($n = 10$), we defined the anterior corona radiata as in Group A but classified it as either 'normal-appearing' (NA; $n = 5$) or 'lesion' ($n = 5$). All ROIs were bilateral. The anterior corona radiata of group A is shown on a directionally encoded color image from DTI and the remaining group A and C ROIs are shown on top of DTI $S_0$ images. The group B ROIs are shown on top of MPRAGE images.

**Table 2** – Diffusion sequences used in the study

| | B-tensor encoding (LTE + STE) | | | Multi-echo (LTE) | |
| --- | --- | --- | --- | --- | --- |
| | Group A | Group B | Group C | Group A | Group C |
| TR [ms] | 7000 | 4000 | 5200 | 7000 | 6900 |
| TE [ms] | 100 | 106 | 106 | 50,90,120,150 | 50,85,120,155 |
| Matrix size | 110×110 | 128×128 | 128×128 | 110×110 | 128×128 |
| Number of slices | 30 | 27 | 35 | 30 | 35 |
| Resolution [mm$^2$] | 2×2 | 2×2 | 2×2 | 2×2 | 2×2 |
| Slice thickness [mm] | 2 | 4 | 4 | 2 | 4 |
| PI factor[a] | 2 | 2 | 2 | 2 | 2 |
| Partial Fourier | 0.75 | 0.75 | 0.75 | 0.75 | 0.75 |
| $b$-values [ms/µm$^2$] | 0.1,0.5,1.0,1.5,2.0[b] | 0.1,0.5,1.0,1.5,2.0[b] | 0.1,0.5,1.0,1.5,2.0[b] | 0.1, 0.5[c] | 0, 0.5[c] |
| $b_\Delta$-values | 1 and 0 | 1 and 0 | 1 and 0 | 1 | 1 |
| # directions / $b$-value | 6, 10, 12, 16, 20 | 6, 6, 10, 16, 30 | 6, 6, 10, 16, 30 | 6, 6 | 1, 6 |
| Number of samples | 128 | 136 | 136 | 12 | 12 |
| Scan time [min] | 15.5 | 9 | 12 | 7 | 4.5 |
| Maxwell compensation[d] | Yes | No | No | Yes | No |
| Head coil array | 64 ch | 20 ch | 20 ch | 64 ch | 20 ch |
| Data prev. published | No | Yes[e] | No | No | No |

[a]GRAPPA; [b]all acquired using both LTE and STE; [c]both acquired using LTE for all TE; [d]Szczepankiewicz et al. (2018); [e]Lampinen et al. (2017);

## 4. Results

*4.1 Brain patterns of microscopic anisotropy from b-tensor encoding*

The representation-based analysis (Eq. 1) produced high-quality maps of the level of microscopic anisotropy ($MK_A$; Fig. 2A). A strong contrast in anisotropy was observed between white and gray matter, with high levels in white matter and low levels in gray matter in general and in the cortex in particular. The level of isotropic heterogeneity ($MK_I$) was generally low, except in regions with substantial partial volume effects between tissue and CSF. The contrast in mean diffusivity (MD) was flat across the brain tissue with similar levels in white and gray matter.

The $MK_A$ values were higher in ROIs expected to feature a higher myelin content (Fig. 3, Table 3). Pure white matter (the anterior corona radiata) featured the highest values of $MK_A$, the thalamus featured intermediate values and the cortex featured the lowest values. Within the thalamus, the $MK_A$ values were significantly higher in the myelin-rich ventrolateral thalamus compared to in the myelin-poor mediodorsal thalamus (mean $\pm$ SD; $0.71 \pm 0.08$ vs. $0.42 \pm 0.05$; $t(4) = 19$, $p < 0.001$). Within the cortex, the $MK_A$ values were significantly higher in the four combined high myelin regions compared to in the four combined low myelin regions ($0.37 \pm 0.04$ vs. $0.20 \pm 0.02$; $t(19) = 31$, $p < 0.001$). Finally, when featuring lesions, the anterior corona radiata exhibited decreased $MK_A$ values and increased values of $MK_I$ and MD, compared to normal-appearing white matter.

There was a correspondence in contrast between the maps of microscopic anisotropy and the myelin-stained histology slice (Fig 2B). Both the $MK_A$ maps and the myelin-stain were bright in white matter and dark in cortical gray matter. The correspondence was seen also between substructures, for example, in the intensity gradient between the myelin-poor mediodorsal thalamus and the myelin-rich ventrolateral thalamus. The $MK_A$ map was comparatively hyperintense in some white matter regions, including the internal capsule and the corpus callous, possibly reflecting a high orientation coherence and anisotropic diffusion in the extracellular space.

*4.2 Biophysical modeling with compartment-specific T2 times*

Precise estimation of compartment-specific T2 times was not possible in the brain (Fig. 4A), which precluded accurate estimation of the 'stick' fraction. The first row shows powder-averaged signal curves in different brain regions (presented by column) with multi-echo signal curves in inset plots. The second row shows how well the minimally constrained model could explain the signal for different fixed 'stick' fractions in terms of the normalized residual variance (Eq. 18). The flat 'valleys' in the residual variance corresponded to acceptable 'stick' fractions between, for example, 0.3 and 0.7 in the anterior corona radiata and between 0.0 and 0.7 in the primary auditory cortex. The cause for the low precision is seen in the signal data. While there was a large effect of varying the b-tensor shape (difference between LTE versus STE reveals microscopic anisotropy), the multi-echo data did not reveal any obvious TE dependence of diffusion (change in slope of signal with TE, Eq. 16). As a result, the data did not support estimation of all parameters of the high-capacity C0 model, and thus wide ranges of 'stick' fractions could explain data equally well in all regions.

The flat 'valleys' were related to wide ranges of acceptable compartment-specific T2 times and diffusivities, shown by the third and fourth rows of figure 4A. For example, acceptable T2 times for the 'stick' compartment ranged between 57 and 93 ms in the anterior corona radiata and between 44 and 118 ms in the primary auditory cortex. Although the isotropic diffusivity was generally higher in the 'ball' compared to the 'stick' compartment, the difference was apparently insufficient to yield precise compartment-specific T2 times (Eq. 17).

Results from white matter lesions stood in contrast to those from the healthy brain. In the lesions, data did allow precise estimation of the 'stick' fraction, and the ranges of data-compatible 'stick' fractions were narrow (Fig. 4B, second row). The signal exhibited a TE dependence (first row) and the 'ball' compartment featured a longer T2 time (third row) and higher isotropic diffusivity (fourth row) compared to the 'stick' compartment. The result is consistent with histological findings in white matter lesions of a loss of tissue integrity that enlarges the extracellular space (Englund and Brun 1990).

Table 4 shows parameter values from all five lesions. The 'stick' fraction exhibited a considerable variation, between approximately 0.25 and 0.45, consistent with white matter lesions lying on a spectrum of tissue changes with varying severity (Gouw et al. 2010). Despite this apparent heterogeneity, the diffusion and relaxation properties of 'sticks' were similar across lesions, and the estimated values were similar to values previously reported in healthy white matter. The 'stick' T2 times were between 60 and 80 ms, similar to in Veraart et al. (2017b) for this region, and the isotropic diffusivities were approximately 0.7 $\mu m^2$/ms, similar to in Veraart et al. (2017b) and Dhital et al. (2017). A possible interpretation of this finding is that the remaining axons inside white matter lesions retain properties close to those in normal tissue.

*4.3 Brain region rankings by the 'stick' fractions of constrained biophysical models*

As expected, all constrained models (Table 1) obtained precise solutions when fit to the combined b-tensor and multi-echo data. Figure 5A exemplifies how the ranges of 'stick' fractions that fit data well (yielded low normalized residual variance, Eq. 18) were narrow (solid lines) compared to for the minimally constrained model (C0; dashed line), and how the different constraints tended to 'prefer' different values of the 'stick' fraction (location of curve minima).

The six constrained models yielded four different rankings of the seven regions with respect to their average 'stick' fractions (Fig. 5B). The ranking obtained with constraint set C1 (as arbitrary reference) was contradicted by a switching of the primary auditory cortex and the mediodorsal thalamus (using C2–C4) and by comparatively high 'stick' fractions in the cortex (using C5) and in white matter lesions (using C6).

No constrained model obtained a good fit both in healthy brain and in white matter lesions (Fig. 5C). All models obtained similar fit qualities in white matter and in the thalamus. In the cortex, however, an overall good fit was only obtained using set C4, which features a CSF compartment. In general, fits were poorer in the primary auditory cortex, compared to in the insular cortex, possibly reflecting differences in partial volume effects with CSF. Using set C5 (as in NODDI), however, yielded a poor fit in both cortical regions despite its CSF compartment, likely due to relating neurite density to mean diffusivity (Lampinen et al. 2017). In white matter lesions, models using the assumption of equal compartment T2 times (Eq. 11; present in C2–C5) yielded comparatively poor fits, consistent with the substantial T2 differences demonstrated in Table 4. The fit within lesions was particularly poor using set C4, likely due to the invalid assumption of equal compartment diffusivities (Table 4).

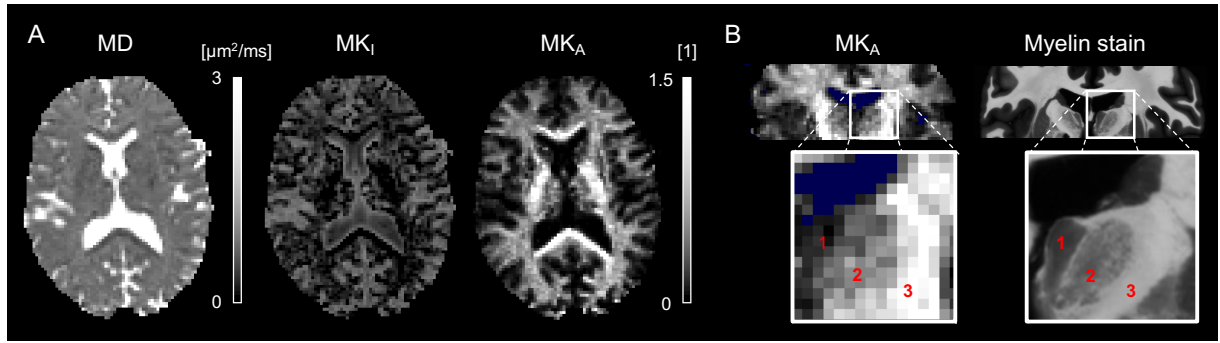

**Figure 2 – Parameter maps from the representation-based analysis (Eq. 1) of b-tensor data.** A) The mean diffusivity (MD) contrast was similar between white and gray matter. The level of isotropic heterogeneity ($MK_I$) was generally low but higher in regions with partial volume effects from CSF. The level of microscopic anisotropy ($MK_A$) was high in white matter and low in gray matter such as the cortex. B) The level of microscopic anisotropy reflected the level of myelination seen in a coronal myelin-stained histology slice (obtained from a different subject). From left to right: (1) the myelin-poor mediodorsal thalamus, (2) the myelin-rich ventrolateral thalamus and (3) the internal capsule. Voxels with MD ≥ 2 $\mu m^2$/ms are colored blue. The shown example is from Group A.

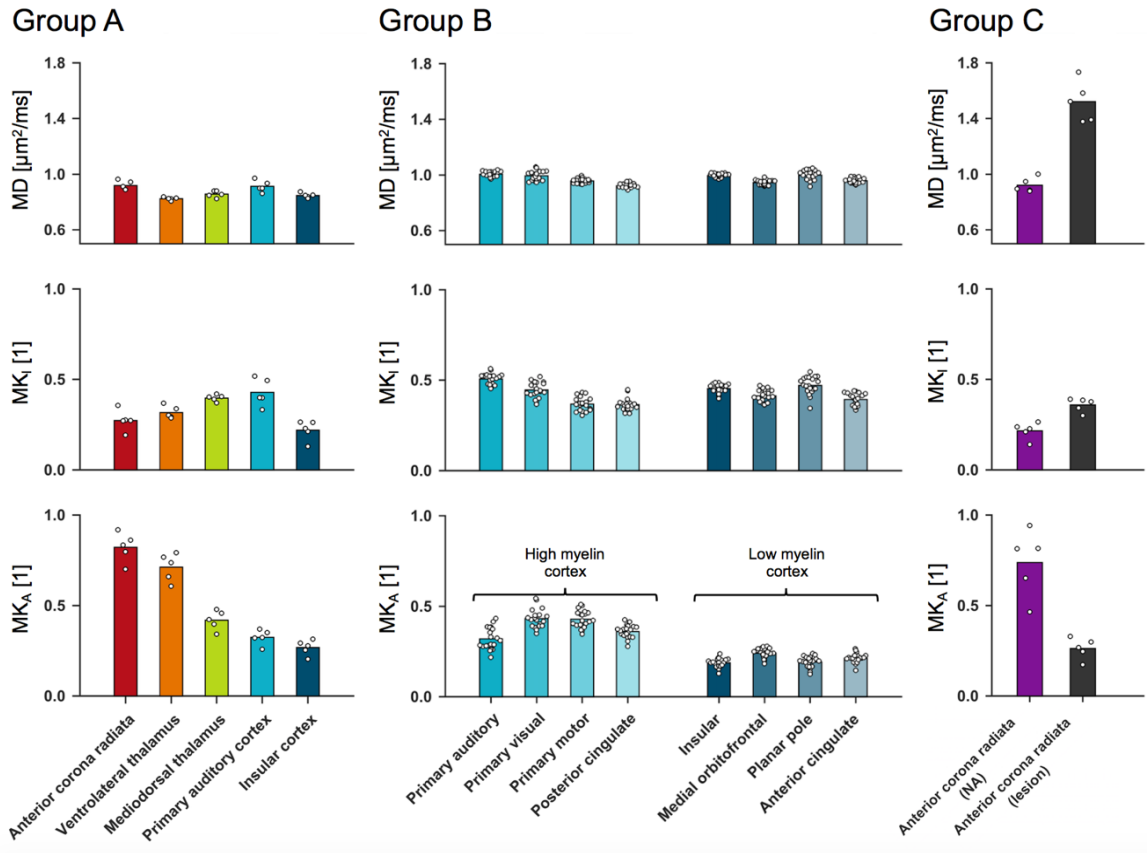

**Figure 3 – Parameter values from the representation-based analysis (Eq. 1) of b-tensor data.** The parameter representing microscopic anisotropy ($MK_A$) yielded the overall strongest contrast between regions, and the $MK_A$ values were higher in regions expected to feature a higher myelin content. In general, the $MK_A$ values were highest in white matter, intermediate in the thalamus and lowest in the cortex. In particular, the $MK_A$ was significantly higher in the myelin-rich ventrolateral thalamus compared to in the more myelin-poor mediodorsal thalamus ($p < 0.001$; group A) and in the combined high myelin cortical regions compared to in the combined low myelin cortical regions in ($p < 0.001$; Group B). Meanwhile, no clear trend was observed in the parameters representing the mean diffusivity (MD) and the isotropic kurtosis ($MK_I$). White matter featuring a lesion (the anterior corona radiata; group C) exhibited reduced $MK_A$ values and increased $MK_I$ and MD values compared to normal-appearing (NA) white matter.

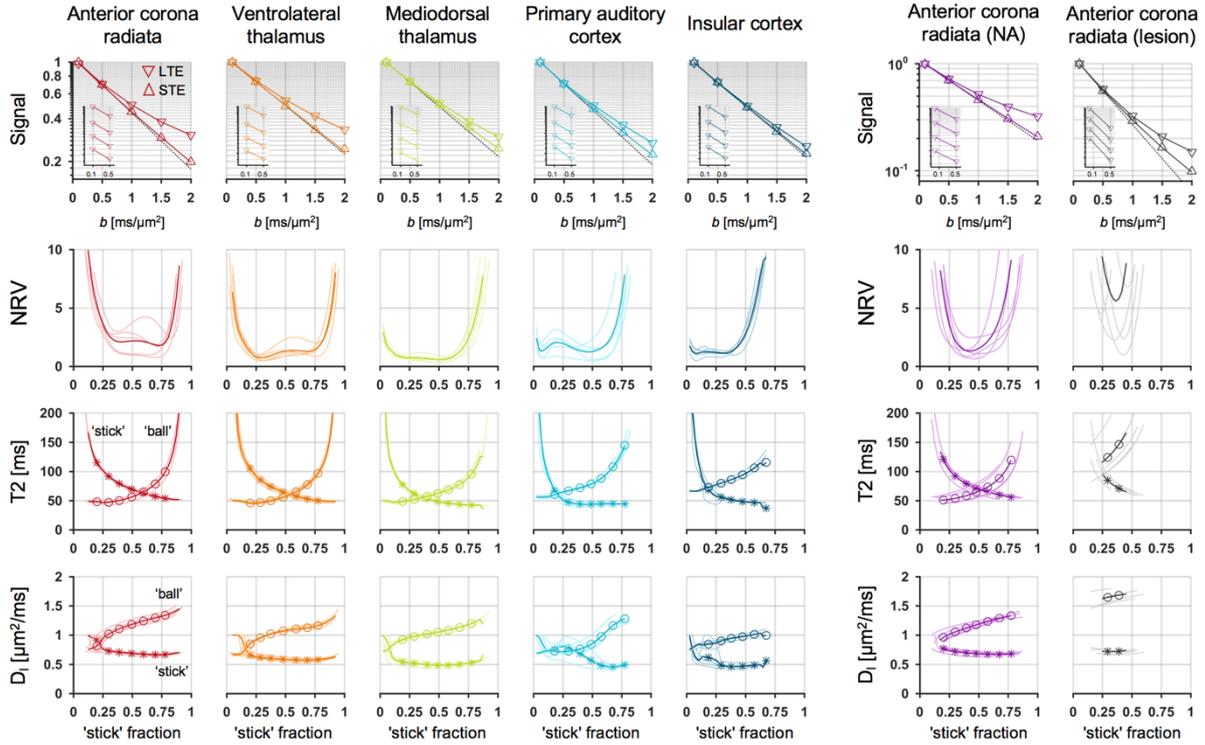

**Figure 4 – Compartment densities could be estimated independently from T2 relaxation in white matter lesions, but not in the healthy brain.** A) In the healthy brain, different regions (sorted by column) exhibited rather different levels of microscopic anisotropy, as seen by the difference between the powder-averaged LTE and STE signal curves (first row). No strong echo-time dependence of diffusion (Eq. 16) was revealed by the slopes of the multi-echo signal curves (inset plots), however, and the regions were not clearly separable by the 'stick' fraction of the minimally constrained model (Eq. 5). The goodness of fit plot (second row; Eq. 18) showed wide ranges of data-compatible (fixed) 'stick' fraction values, reflecting wide ranges of data-compatible compartment-specific T2 times (third row). For most 'stick' fractions, however, the isotropic diffusivity was higher in the 'ball' compartment compared to in the 'stick' compartment (fourth row). B) In white matter (anterior corona radiata) with a lesion, precise estimation of the 'stick' fraction together with compartment-specific T2 times was possible. The 'ball' compartment featured a longer T2 time and a considerably higher isotropic diffusivity compared to the 'stick' compartment, resulting in a sufficient echo-time dependence of diffusion to yield a narrow range of data-compatible 'stick' fractions. The plots use shaded lines for individual subjects and thick lines for group-averages.

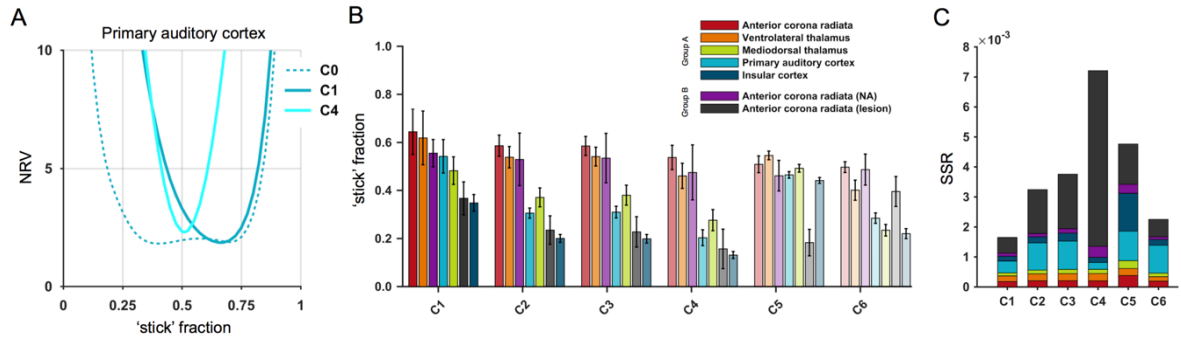

**Figure 5 – Using different constraints yielded different interpretations of the 'stick' fraction.** A) The constrained models (solid lines, here C1 and C4; Table 1) yielded precise estimates of the 'stick' fraction, seen by the narrow ranges of (fixed) values that yielded a high goodness of fit (Eq. 18) compared to for the minimally constrained model (dashed line, C0). B) The six constrained models yielded four different rankings of seven regions with respect to the 'stick' fraction. Thus, most of these 'stick' fractions lacked ordinal accuracy across the domain represented by the healthy brain and white matter lesions, and using them as indices of the neurite density yields constraint-dependent results. For example, constraint sets C1 and C6 would indicate a lower neurite density in the primary auditory cortex compared to in the mediodorsal thalamus while sets C2–C4 would indicate the opposite. C) The quality of fit (in terms of a low sum of squared residuals, SSR) was generally high in white matter and in the thalamus but not always in the cortex or in white matter lesions. In the cortex, only set C4, which featured a CSF compartment, obtained a good fit. In white matter lesions, only sets allowing different compartment T2 times (C1 and C6) obtained a good fit. Note that quality of fit cannot necessarily be used to test for poor constraints. Sets C1 and C6 yielded conflicting results when comparing the ventrolateral thalamus and the anterior corona radiata despite a good fit quality, and all comparisons using sets C3 and C4 were identical despite very different fit qualities in the cortex.

**Table 3** – Parameter values from the representation-based analysis (Eq. 1) of b-tensor data (Fig.1 ROIs), with means and inter-subject standard deviations.

| Group A | MD [$\mu m^2$/ms] | MKI [1] | MKA [1] |
|---|---|---|---|
| Anterior corona radiata | 0.92 (0.03) | 0.27 (0.06) | 0.82 (0.08) |
| Ventrolateral thalamus | 0.82 (0.01) | 0.32 (0.03) | 0.71 (0.08) |
| Mediodorsal thalamus | 0.86 (0.02) | 0.40 (0.02) | 0.42 (0.05) |
| Primary auditory cortex | 0.91 (0.04) | 0.43 (0.08) | 0.32 (0.04) |
| Insular cortex | 0.85 (0.02) | 0.22 (0.05) | 0.27 (0.04) |

| Group B (cortex) | MD [$\mu m^2$/ms] | MKI [1] | MKA [1] |
|---|---|---|---|
| Primary auditory | 1.01 (0.02) | 0.51 (0.03) | 0.32 (0.06) |
| Primary visual | 0.99 (0.03) | 0.45 (0.04) | 0.43 (0.05) |
| Primary motor | 0.96 (0.02) | 0.38 (0.04) | 0.43 (0.05) |
| Posterior cingulate | 0.92 (0.02) | 0.37 (0.03) | 0.36 (0.04) |
| Insular | 1.00 (0.01) | 0.45 (0.02) | 0.19 (0.02) |
| Medial orbitofrontal | 0.95 (0.02) | 0.41 (0.03) | 0.24 (0.03) |
| Polar plane | 1.00 (0.03) | 0.47 (0.05) | 0.19 (0.03) |
| Anterior cingulate | 0.96 (0.02) | 0.39 (0.03) | 0.21 (0.03) |

| Group C (Anterior corona radiata) | MD [$\mu m^2$/ms] | MKI [1] | MKA [1] |
|---|---|---|---|
| Normal-appearing | 0.92 (0.05) | 0.22 (0.05) | 0.74 (0.18) |
| White matter lesion | 1.52 (0.15) | 0.36 (0.04) | 0.26 (0.06) |

**Table 4 –** Best-fitting solutions for the minimally constrained model in the five white matter lesion ROIs (Fig. 4B).

| $f_S$ | T2$_B$ | T2$_S$ | $D_{I;B}$ | $D_{I;S}$ |
|-------|--------|--------|-----------|-----------|
| 0.46  | 173    | 63     | 1.73      | 0.71      |
| 0.43  | 137    | 67     | 1.72      | 0.71      |
| 0.43  | 123    | 70     | 1.63      | 0.67      |
| 0.35  | 110    | 84     | 1.56      | 0.74      |
| 0.24  | 169    | 83     | 1.78      | 0.75      |

Relaxation times are in ms and diffusivities are in $\mu m^2/ms$.

## 5. Discussion

Mapping neurite density requires a correct model for diffusion in both axons and dendrites. The conventional neurite assumption is that these structures exhibit highly anisotropic ('stick-like') diffusion. If this is the case, the respective prevalence of these structures should contribute similarly to microscopic diffusion anisotropy in the brain, and we would expect a contrast in $MK_A$ that bears similarity to the contrast in neurite density. In cortical gray matter, histology studies in mice have reported a rather even split between axons and dendrites and a neurite volume fraction of approximately 60% (Ikari et al. 1981, Chklovskii et al. 2002, Braitenberg et al. 2013). In white matter, histology studies have reported a neurite density (intra-axonal volume fraction) of 30–50% in the corpus callosum of macaque (Stikov et al. 2015) and mice (Jelescu et al. 2016b) and in the rat spinal cord (Xu et al. 2014). Together with a 20–50% volume fraction of myelin water (Xu et al. 2014, Stikov et al. 2015, Jelescu et al. 2016b) with short T2 time (Mackay et al. 1994), an axonal volume fraction of 30–50% corresponds to an *MR-visible* neurite signal fraction, $v_{axons} / (1 - v_{myelin})$, in the vicinity of the approximate 60% reported for the cortex. Thus, histology suggests that the signal fraction of neurites as observed with dMRI should be similar between gray and white matter, wherefore, by the neurite assumption, the respective levels of microscopic diffusion anisotropy should be similar. In contrast, however, we found that microscopic anisotropy differs substantially between gray and white matter (Fig. 2–3, Table 3). Furthermore, regional differences were associated with myelin content, and thus axons, both qualitatively (Fig. 2B) and quantitatively (Fig. 3, Table 3), consistent with previous observations in fixed tissue by Jespersen et al (2010). These observations contradict the neurite assumption and suggests that highly anisotropic diffusion is a feature of axons but not dendrites. A non-zero radial diffusivity in dendrites could be due to, for example, fast exchange with the extracellular space or between short dendritic segments with different orientations. A low axial diffusivity in dendrites is not a sufficient cause, however, since it would have manifested in a higher $MK_I$, and constraints that allow an independent 'stick' diffusivity (such as C2) should have yielded an 'adjusted' 'stick' fraction (Table 1; Fig. 5B). Our observations could potentially be explained, however, by a comparatively short neurite T2 time in gray matter, since that would confound a T2-weighted metric such as the $MK_A$. Notably, such differences would also confound volume fraction estimates under the density assumption, leading to the second part of our investigation.

Mapping compartment densities using dMRI with a single TE requires that T2 times are approximately equal across tissue components. In white matter lesions, our analysis of b-tensor and multi-echo data found that water with isotropic diffusivity exhibited a longer T2 time compared to water exhibiting 'stick-like' diffusion (Fig. 4B, Table 4). In the healthy brain, however, compartment-specific T2 times could not be estimated (Fig. 4A) and potential confounding effects on the 'stick' fraction could not be determined. This lack of precision also prevented us from directly assessing whether a short T2 time of neurites may explain the low levels of microscopic anisotropy in gray matter. However, the results did indicate the approximate ranges of compartment T2 times and diffusivities for which data are compatible with this hypothesis. Table 5 shows group-averaged parameter values for the solutions in the flat 'valleys' from figure 4A. In short, the 'stick' fraction could assume values in a vicinity of the expected neurite density (0.5–0.7; gray area) if the T2 time difference between 'neurite' and 'non-neurite' tissue was small in white matter but large in gray matter. For example, ballpark figures would indicate similar T2 times for neurite and non-neurite tissue (60–80 ms) in the anterior corona radiata, but much shorter T2 times for neurites (40–50 ms) compared to non-neurite tissue (80–120 ms) in the cortex. The solutions would also imply a higher axial diffusivity of neurites in white matter (approximately 2.1 $\mu m^2$/ms in the anterior corona radiata) compared to in gray matter (approximately 1.5 $\mu m^2$/ms in the cortex). Thus, estimating neurite density may require correcting for such differences, which is a difficult task even using multi-shell multi-b-tensor and multi-echo data (Fig. 4A). Note that application of the density assumption of equal compartment T2 times yields the heterogenous anisotropy observed in the representation-based analysis (Figs. 2 and 3; Table 3), and an interpretation that is in apparent conflict with histology. These results suggest that estimating neurite density from conventional single-echo dMRI has low accuracy. Whether such parameters could still be indices for neurite density was explored in the third part of our investigation.

An index with sensitivity to a physical quantity could be useful even if lacking absolute accuracy, for example, by having the ordinal accuracy to allow comparisons between regions, conditions, or time points. When confounding effects grow large compared to the effect of interest, however, results should become increasingly dependent on the choice of model constraints. Here, when ranking seven regions with respect to the 'stick' fraction, using six different sets of constraints yielded four different rankings (Fig. 5B). From this, we draw two conclusions concerning the range of conditions represented by the healthy brain and white matter lesions. First, as there can only be one true ranking with respect to a given quantity, such as neurite density, this range of conditions violates the domain of validity for most of these 'stick' fractions. Second, this range of conditions likely features important confounding factors. We suspect compartmental T2 differences to be one such factor, both because of its strong impact on volume fractions (Fig. 4), and because constraining it in three different ways (C1: Eq. 10; C2–C5: Eq. 11; C6: Eq. 14) yielded three different rankings (Fig. 5B). Within the groups of models using the density assumption (Eq. 11), the disagreement between set C5 and the others could be due to its treatment of isotropic heterogeneity, a confounding factor for microscopic anisotropy (Mitra 1995), which results in incompatibility with STE data (Lampinen et al. 2017). Note that invalid constraints do not necessarily cause low goodness of fit, seeing that sets C1 and C6 yielded different rankings between the ventrolateral thalamus and the anterior corona radiata (NA; Fig. 5B) where this was not an issue (Fig. 5C). Although this experiment was discouraging for the prospect of finding a widely applicable neurite density index, overcoming some of these issues should be possible through more extensive acquisitions and/or additional validation work.

Our investigation indicates that estimating the neurite density accurately requires accounting for different diffusion and/or T2 properties between axons and dendrites. As a consequence, true neurite density estimation may not be feasible with moderate b-values and limited ranges of TE. Obtaining indices of the neurite (or axonal) density may be possible, although their usefulness is limited to conditions where confounding factors are small or can be reliably constrained. One such condition could be healthy white matter, where the effects of T2 relaxation (Beaulieu et al. 1998, Clark and Le Bihan 2000) and isotropic heterogeneity (Dhital et al. 2015, Szczepankiewicz et al. 2015) appear to be small, and where excellent (ex-vivo) correlations have been demonstrated between the 'stick' fraction and histological metrics of axonal density (Jespersen et al. 2010). Thus, in healthy white matter, existing methods such as WMTI (Fieremans et al. 2011), NODDI or SMT may be able to capture the relative variation of axonal density. However, the often used 'neurite' label suggests a domain of validity that also encompasses gray matter, where neurites may exhibit different diffusion and/or relaxation properties (Table 3) and their disentangling may be further complicated by CSF contamination. Extending a neurite density index to examine pathology entails additional challenges, including new tissue components in tumors (Wen et al. 2015), exchange between tissue components in stroke (Lätt et al. 2009), and perturbing effects on T2 relaxation (Fig. 4B) and/or demyelination in degenerative conditions, where the latter could mimic axonal loss simply by expanding the pool of MR-visible water with low diffusion anisotropy. In the future search for the neurite density with dMRI, studies should attempt to elucidate the contribution of dendrites to microscopic anisotropy, to inform constraints by exploring relations between diffusion and non-diffusion properties, and to delineate the domains of validity for prospective index parameters. The current state of evidence suggests that reporting results in terms of the 'stick' signal fraction, or possible the axonal density index in healthy white matter, would promote the most useful interpretations of clinical studies using biophysical models.

We note several limitations with the present study. First, the waveforms used to produce linear and spherical tensor encoding had slightly different timing properties. Lundell et al. (2017) demonstrated in ex vivo monkey brain that using waveforms with different 'spectral content' may confuse time-dependent diffusion for microscopic anisotropy. We investigated this by comparing LTE obtained using waveforms that have either 'tuned' or 'detuned' encoding spectra (e.g. similar or different timing properties) with respect to the waveform used for STE (supporting information). No systematic effects of diffusion time were observed, which is in line with multiple studies showing negligible time dependence in living brain tissue for clinically relevant diffusion times, both in humans (Clark and Le Bihan 2000, Clark et al. 2001, Nilsson et al. 2009) and animals (Niendorf et al. 1996, Ronen et al. 2006). Second, the waveforms used for groups B and C were not optimized for negligible concomitant fields, which may induce a positive bias in microscopic anisotropy (Szczepankiewicz et al. 2018). From assessing the data for characteristic gross signal errors, however, we do not believe that the effect was large for the waveforms applied in this study. Furthermore, such bias should have no systematic impact on the inter-cortical comparisons performed with the group B ROIs; and the group C data was only extracted from relatively deep parts of the brain (the corona radiata) where the effects of concomitant fields should be small (Szczepankiewicz and Nilsson 2018). Third, due to the SNR penalty from studying deep gray matter while using small voxels, we acquired multi-echo data for a maximum b-value of 0.5 ms/$\mu$m$^2$. Accordingly, we could only rely on a TE dependence of the mean diffusivity to separate compartment T2 times (Eq. 16), an effect that has previously been demonstrated to be small in the healthy brain (Beaulieu et al. 1998, Clark and Le Bihan 2000, Qin et al. 2009, Tax et al. 2017, Veraart et al. 2017b). Acquiring multi-echo data for higher b-values may be a promising way forward (Veraart et al. 2017b), and our simulations indicate that extending our multi-echo dataset to include higher b-values could yield sufficient precision in most of the studied regions (supporting information). Fourth, our minimally constrained model featured the 'ball' constraint of isotropic diffusion outside of 'stick-like' structures, which may be inaccurate in regions with high orientation coherence (Fieremans et al. 2011, Reisert et al. 2017, Novikov et al. 2018b). As a consequence, our investigation of compartment-specific T2 times in white matter was limited to the highly disperse anterior corona radiata (Mollink et al. 2017). Fifth, our minimally constrained model fixed the T1 relaxation and proton density properties to be equal across tissue compartments. Although we believe that these effects may be important and should be considered in future studies, they were outside the scope of the current work. Finally, we acknowledge that more work is needed to elucidate the relative contribution from dendrites to microscopic anisotropy, as well as the role of exchange.

**Table 5** – Data-compatible 'stick' fractions ($f_S$) for the minimally constrained model in the healthy brain (Fig. 4A), with corresponding parameters.

| | $f_S$ | 0.1 | 0.2 | 0.3 | 0.4 | 0.5 | 0.6 | 0.7 | 0.8 | 0.9 |
|---|---|---|---|---|---|---|---|---|---|---|
| **Anterior corona radiata** | T2$_B$ | 49 | 48 | 47 | 51 | 56 | 65 | 80 | 107 | 214 |
| | T2$_S$ | 211 | 115 | 92 | 78 | 69 | 62 | 57 | 53 | 52 |
| | $D_{I;B}$ | 0.75 | 0.81 | 1.02 | 1.12 | 1.19 | 1.25 | 1.30 | 1.35 | 1.45 |
| | $D_{I;S}$ | 1.00 | 0.91 | 0.73 | 0.70 | 0.69 | 0.67 | 0.66 | 0.67 | 0.70 |
| **Ventrolateral thalamus** | T2$_B$ | 50 | 47 | 47 | 51 | 57 | 66 | 80 | 105 | 195 |
| | T2$_S$ | 164 | 104 | 84 | 72 | 63 | 57 | 52 | 50 | 49 |
| | $D_{I;B}$ | 0.67 | 0.83 | 0.96 | 1.03 | 1.08 | 1.11 | 1.13 | 1.18 | 1.29 |
| | $D_{I;S}$ | 1.00 | 0.71 | 0.62 | 0.60 | 0.59 | 0.58 | 0.57 | 0.59 | 0.62 |
| **Mediodorsal thalamus** | T2$_B$ | 48 | 49 | 53 | 57 | 63 | 70 | 81 | 103 | 182 |
| | T2$_S$ | 111 | 76 | 61 | 53 | 48 | 45 | 43 | 42 | 43 |
| | $D_{I;B}$ | 0.82 | 0.93 | 0.98 | 1.02 | 1.05 | 1.09 | 1.14 | 1.22 | 1.34 |
| | $D_{I;S}$ | 0.63 | 0.54 | 0.51 | 0.49 | 0.49 | 0.49 | 0.51 | 0.54 | 0.60 |
| **Primary auditory ctx** | T2$_B$ | 56 | 62 | 67 | 72 | 80 | 92 | 115 | 162 | 529 |
| | T2$_S$ | 108 | 60 | 47 | 44 | 44 | 45 | 45 | 44.6 | 45.9 |
| | $D_{I;B}$ | 0.72 | 0.72 | 0.72 | 0.77 | 0.86 | 1.02 | 1.20 | 1.31 | 1.46 |
| | $D_{I;S}$ | 0.77 | 0.78 | 0.78 | 0.68 | 0.56 | 0.47 | 0.46 | 0.51 | 0.59 |
| **Insular cortex** | T2$_B$ | 67 | 74 | 80 | 87.5 | 97 | 110 | 120 | 172 | 716 |
| | T2$_S$ | 114 | 64 | 55 | 50.7 | 48 | 47 | 37 | 39 | 41 |
| | $D_{I;B}$ | 0.84 | 0.85 | 0.90 | 0.95 | 0.98 | 1.02 | 1.00 | 1.07 | 1.19 |
| | $D_{I;S}$ | 0.58 | 0.61 | 0.49 | 0.46 | 0.47 | 0.49 | 0.49 | 0.65 | 0.70 |

Shown values are group-averages from the flat 'valleys' of (best fitting) solutions in Fig. 4A, with remaining values shaded. The gray area corresponds to 'stick' fractions similar to approximate neurite density estimates from histology (0.5–0.7). Relaxation times are in ms and diffusivities are in $\mu m^2$/ms.

# 6. Conclusions

This work investigated three challenges associated with estimating the neurite density with dMRI. Our results suggest that neurite density estimation requires accounting for different diffusion and/or T2 properties between axons and dendrites, which is not feasible with conventional single-echo multi-shell dMRI. Obtaining parameters that index the neurite (or axonal) density should be possible, but confounding factors may prevent their use outside healthy white matter. Finding a reliable neurite density index in gray matter or pathology likely requires finding correlations between diffusion and non-diffusion properties as well as an improved understanding of diffusion in dendrites.

## Acknowledgements

The authors acknowledge the Swedish Research Council (grant no. 2016-03443), the Swedish Foundation for Strategic Research (grant no. AM13-0090), Random Walk Imaging (grant no MN15), NIH grants R01MH074794, P41EB015902, P41EB015898. The Linnaeus Environment Thinking in Time: Cognition, Communication and Learning, financed by the Swedish Research Council (349-2007-869). We thank Siemens Healthineers for providing access to the pulse sequence programming environment.

## Conflicts of interest

MN declares research support from Random Walk Imaging, and patent applications in Sweden (1250453-6 and 1250452-8), USA (61/642 594 and 61/642 589), and PCT (SE2013/050492 and SE2013/050493). Remaining authors declare no conflict of interest.

# REFERENCES

Alexander, D. C., T. B. Dyrby, M. Nilsson and H. Zhang (2017). "Imaging brain microstructure with diffusion MRI: Practicality and applications." NMR in Biomedicine.

Assaf, Y. and P. J. Basser (2005). "Composite hindered and restricted model of diffusion (CHARMED) MR imaging of the human brain." Neuroimage **27**(1): 48-58.

Ballabh, P., A. Braun and M. J. N. o. d. Nedergaard (2004). "The blood–brain barrier: an overview: structure, regulation, and clinical implications." **16**(1): 1-13.

Basser, P. J., J. Mattiello and D. LeBihan (1994). "MR diffusion tensor spectroscopy and imaging." Biophysical journal **66**(1): 259.

Beaulieu, C., F. R. Fenrich and P. S. Allen (1998). "Multicomponent water proton transverse relaxation and T2-discriminated water diffusion in myelinated and nonmyelinated nerve." Magnetic resonance imaging **16**(10): 1201-1210.

Behrens, T. E., M. W. Woolrich, M. Jenkinson, H. Johansen-Berg, R. G. Nunes, S. Clare, P. M. Matthews, J. M. Brady and S. M. Smith (2003). "Characterization and propagation of uncertainty in diffusion-weighted MR imaging." Magnetic resonance in medicine **50**(5): 1077-1088.

Bock, N. A., A. Kocharyan, J. V. Liu and A. C. Silva (2009). "Visualizing the entire cortical myelination pattern in marmosets with magnetic resonance imaging." Journal of neuroscience methods **185**(1): 15-22.

Braitenberg, V. and A. Schüz (2013). Cortex: statistics and geometry of neuronal connectivity, Springer Science & Business Media.

Callaghan, P., K. Jolley and J. Lelievre (1979). "Diffusion of water in the endosperm tissue of wheat grains as studied by pulsed field gradient nuclear magnetic resonance." Biophysical journal **28**(1): 133-141.

Chenevert, T. L., L. D. Stegman, J. M. Taylor, P. L. Robertson, H. S. Greenberg, A. Rehemtulla and B. D. Ross (2000). "Diffusion magnetic resonance imaging: an early surrogate marker of therapeutic efficacy in brain tumors." Journal of the National Cancer Institute **92**(24): 2029-2036.

Chklovskii, D. B., T. Schikorski and C. F. Stevens (2002). "Wiring optimization in cortical circuits." Neuron **34**(3): 341-347.

Clark, C. A., M. Hedehus and M. E. Moseley (2001). "Diffusion time dependence of the apparent diffusion tensor in healthy human brain and white matter disease." Magnetic resonance in medicine **45**(6): 1126-1129.

Clark, C. A. and D. Le Bihan (2000). "Water diffusion compartmentation and anisotropy at high b values in the human brain." Magnetic Resonance in Medicine **44**(6): 852-859.

Danos, P., B. Baumann, A. Krämer, H.-G. Bernstein, R. Stauch, D. Krell, P. Falkai and B. Bogerts (2003). "Volumes of association thalamic nuclei in schizophrenia: a postmortem study." Schizophrenia Research **60**(2): 141-155.

de Almeida Martins, J. P. and D. J. S. r. Topgaard (2018). "Multidimensional correlation of nuclear relaxation rates and diffusion tensors for model-free investigations of heterogeneous anisotropic porous materials." **8**(1): 2488.

De Santis, S., D. Barazany, D. K. Jones and Y. Assaf (2016). "Resolving relaxometry and diffusion properties within the same voxel in the presence of crossing fibres by combining inversion recovery and diffusion-weighted acquisitions." Magnetic resonance in medicine **75**(1): 372-380.

Dhital, B., E. Kellner, M. Reisert and V. G. Kiselev (2015). "Isotropic diffusion weighting provides insight on diffusion compartments in human brain white matter in vivo. In Proceedings of the 23rd Annual Meeting of ISMRM, Toronto, Canada, 2015. p. 2788.".

Dhital, B., M. Reisert, E. Kellner and V. G. Kiselev (2017). "Intra-axonal Diffusivity in Brain White Matter." arXiv preprint arXiv:1712.04565.

Douaud, G., S. Jbabdi, T. E. Behrens, R. A. Menke, A. Gass, A. U. Monsch, A. Rao, B. Whitcher, G. Kindlmann and P. M. Matthews (2011). "DTI measures in crossing-fibre areas: increased diffusion anisotropy reveals early white matter alteration in MCI and mild Alzheimer's disease." Neuroimage **55**(3): 880-890.

Englund, E. and A. Brun (1990). "White matter changes in dementia of Alzheimer's type: the difference in vulnerability between cell compartments." Histopathology **16**(5): 433-439.

Englund, E., A. Brun and B. Persson (1987). "Correlations between histopathologic white matter changes and proton MR relaxation times in dementia." Alzheimer disease and associated disorders **1**(3): 156-170.

Eriksson, S., S. Lasič, M. Nilsson, C.-F. Westin and D. Topgaard (2015). "NMR diffusion-encoding with axial symmetry and variable anisotropy: Distinguishing between prolate and oblate microscopic diffusion tensors with unknown orientation distribution." The Journal of chemical physics **142**(10): 104201.

Eriksson, S., S. Lasic and D. Topgaard (2013). "Isotropic diffusion weighting in PGSE NMR by magic-angle spinning of the q-vector." Journal of Magnetic Resonance **226**: 13-18.

Fieremans, E., J. H. Jensen and J. A. Helpern (2011). "White matter characterization with diffusional kurtosis imaging." Neuroimage **58**(1): 177-188.

Goodfellow, I., Y. Bengio and A. Courville (2016). "Deep Learning." MIT Press.

Gouw, A. A., A. Seewann, W. M. Van Der Flier, F. Barkhof, A. M. Rozemuller, P. Scheltens and J. J. Geurts (2010). "Heterogeneity of small vessel disease: a systematic review of MRI and histopathology correlations." Journal of Neurology, Neurosurgery & Psychiatry: jnnp. 2009.204685.

Hopf, A. (1955). "Über die Verteilung myeloarchitektonischer Merkmale in der isokortikalen Schläfenlappenrinde beim Menschen." J Hirnforsch **2**(1): 36-54.

Hopf, A. (1956). "Über die Verteilung myeloarchitektonischer Merkmale in der Stirnhirnrinde beim Menschen." J Hirnforsch **2**(4): 311-333.

Hopkins, A., H. Yeung and C. Bratton (1986). "Multiple field strength in vivo T1 and T2 for cerebrospinal fluid protons." Magnetic resonance in medicine **3**(2): 303-311.

Horsfield, M. A. and D. K. Jones (2002). "Applications of diffusion-weighted and diffusion tensor MRI to white matter diseases–a review." NMR in Biomedicine **15**(7-8): 570-577.

Ikari, K. and M. Hayashi (1981). "Aging in the Neuropil of Cerebral Cortex–A Quantitative Ultrastructural Study." Psychiatry and Clinical Neurosciences **35**(4): 477-486.

Jelescu, I. O., J. Veraart, E. Fieremans and D. S. Novikov (2016a). "Degeneracy in model parameter estimation for multi-compartmental diffusion in neuronal tissue." NMR in Biomedicine **29**(1): 33-47.

Jelescu, I. O., M. Zurek, K. V. Winters, J. Veraart, A. Rajaratnam, N. S. Kim, J. S. Babb, T. M. Shepherd, D. S. Novikov and S. G. Kim (2016b). "In vivo quantification of demyelination and recovery using compartment-specific diffusion MRI metrics validated by electron microscopy." Neuroimage **132**: 104-114.

Jensen, J. H., J. A. Helpern, A. Ramani, H. Lu and K. Kaczynski (2005). "Diffusional kurtosis imaging: The quantification of non-gaussian water diffusion by means of magnetic resonance imaging." Magnetic Resonance in Medicine **53**(6): 1432-1440.

Jespersen, S. N., C. R. Bjarkam, J. R. Nyengaard, M. M. Chakravarty, B. Hansen, T. Vosegaard, L. Østergaard, D. Yablonskiy, N. C. Nielsen and P. Vestergaard-Poulsen (2010). "Neurite density from magnetic resonance diffusion measurements at ultrahigh field: comparison with light microscopy and electron microscopy." Neuroimage **49**(1): 205-216.

Jespersen, S. N., C. D. Kroenke, L. Østergaard, J. J. Ackerman and D. A. Yablonskiy (2007). "Modeling dendrite density from magnetic resonance diffusion measurements." Neuroimage **34**(4): 1473-1486.

Jespersen, S. N., H. Lundell, C. K. Sønderby and T. B. Dyrby (2013). "Orientationally invariant metrics of apparent compartment eccentricity from double pulsed field gradient diffusion experiments." NMR in Biomedicine **26**(12): 1647-1662.

Kaden, E., N. D. Kelm, R. P. Carson, M. D. Does and D. C. Alexander (2016). "Multi-compartment microscopic diffusion imaging." NeuroImage **139**: 346-359.

Kaden, E., T. R. Knösche and A. J. N. Anwander (2007). "Parametric spherical deconvolution: inferring anatomical connectivity using diffusion MR imaging." **37**(2): 474-488.

Kiselev, V. G. (2017). "Fundamentals of diffusion MRI physics." NMR in Biomedicine **30**(3).

Kiselev, V. G. and K. A. Il'yasov (2007). "Is the "biexponential diffusion" biexponential?" Magnetic resonance in medicine **57**(3): 464-469.

Klein, S., M. Staring, K. Murphy, M. Viergever and J. P. Pluim (2010). "Elastix: a toolbox for intensity-based medical image registration." Medical Imaging, IEEE Transactions on **29**(1): 196-205.

Kroenke, C. D., J. J. Ackerman and D. A. Yablonskiy (2004). "On the nature of the NAA diffusion attenuated MR signal in the central nervous system." Magnetic resonance in medicine **52**(5): 1052-1059.

Lampinen, B., F. Szczepankiewicz, J. Mårtensson, D. van Westen, P. C. Sundgren and M. Nilsson (2017). "Neurite density imaging versus imaging of microscopic anisotropy in diffusion MRI: A model comparison using spherical tensor encoding." NeuroImage **147**: 517-531.

Lasič, S., F. Szczepankiewicz, S. Eriksson, M. Nilsson and D. Topgaard (2014). "Microanisotropy imaging: quantification of microscopic diffusion anisotropy and orientational order parameter by diffusion MRI with magic-angle spinning of the q-vector." Frontiers in Physics **2**: 11.

Lätt, J., M. Nilsson, D. van Westen, R. Wirestam, F. Ståhlberg and S. Brockstedt (2009). "Diffusion-weighted MRI measurements on stroke patients reveal water-exchange mechanisms in sub-acute ischaemic lesions." NMR in Biomedicine **22**(6): 619-628.

Lawrenz, M. and J. Finsterbusch (2013). "Double-wave-vector diffusion-weighted imaging reveals microscopic diffusion anisotropy in the living human brain." Magnetic resonance in medicine **69**(4): 1072-1082.

Lawrenz, M. and J. J. M. r. i. m. Finsterbusch (2018). "Detection of microscopic diffusion anisotropy in human cortical gray matter in vivo with double diffusion encoding."

Le Bihan, D., E. Breton, D. Lallemand, P. Grenier, E. Cabanis and M. Laval-Jeantet (1986). "MR imaging of intravoxel incoherent motions: application to diffusion and perfusion in neurologic disorders." Radiology **161**(2): 401-407.

Lebel, C., L. Walker, A. Leemans, L. Phillips and C. Beaulieu (2008). "Microstructural maturation of the human brain from childhood to adulthood." Neuroimage **40**(3): 1044-1055.

Löbel, U., J. Sedlacik, D. Güllmar, W. A. Kaiser, J. R. Reichenbach and H.-J. Mentzel (2009). "Diffusion tensor imaging: the normal evolution of ADC, RA, FA, and eigenvalues studied in multiple anatomical regions of the brain." Neuroradiology **51**(4): 253-263.

Lundell, H., M. Nilsson, T. B. Dyrby, G. J. Parker, P. L. H. Cristinacce, F. Zhou, D. Topgaard and S. Lasic (2017). "Microscopic anisotropy with spectrally modulated q-space trajectory encoding. In proceedings for the 25th Annual Meeting of ISMRM, Honolulu, Hawaii, USA, 2017. p. 1086.".

Mackay, A., K. Whittall, J. Adler, D. Li, D. Paty and D. Graeb (1994). "In vivo visualization of myelin water in brain by magnetic resonance." Magnetic Resonance in Medicine **31**(6): 673-677.

McKinnon, E. T., J. H. Jensen, G. R. Glenn and J. A. J. M. r. i. Helpern (2017). "Dependence on b-value of the direction-averaged diffusion-weighted imaging signal in brain." **36**: 121-127.

Mitra, P. P. (1995). "Multiple wave-vector extensions of the NMR pulsed-field-gradient spin-echo diffusion measurement." Physical Review B **51**(21): 15074.

Mollink, J., M. Kleinnijenhuis, A.-M. v. C. van Walsum, S. N. Sotiropoulos, M. Cottaar, C. Mirfin, M. P. Heinrich, M. Jenkinson, M. Pallebage-Gamarallage and O. Ansorge (2017). "Evaluating fibre orientation dispersion in white matter: Comparison of diffusion MRI, histology and polarized light imaging." Neuroimage **157**: 561-574.

Morris, R., G. Paxinos and M. Petrides (2000). "Architectonic analysis of the human retrosplenial cortex." Journal of Comparative Neurology **421**(1): 14-28.

Moseley, M., Y. Cohen, J. Mintorovitch, L. Chileuitt, H. Shimizu, J. Kucharczyk, M. Wendland and P. Weinstein (1990). "Early detection of regional cerebral ischemia in cats: comparison of diffusion-and T2-weighted MRI and spectroscopy." Magnetic resonance in medicine **14**(2): 330-346.

Niendorf, T., R. M. Dijkhuizen, D. G. Norris, M. van Lookeren Campagne and K. Nicolay (1996). "Biexponential diffusion attenuation in various states of brain tissue: implications for diffusion-weighted imaging." Magnetic Resonance in Medicine **36**(6): 847-857.

Nilsson, M., S. Lasič, I. Drobnjak, D. Topgaard and C. F. Westin (2017). "Resolution limit of cylinder diameter estimation by diffusion MRI: The impact of gradient waveform and orientation dispersion." NMR in Biomedicine.

Nilsson, M., J. Lätt, E. Nordh, R. Wirestam, F. Ståhlberg and S. Brockstedt (2009). "On the effects of a varied diffusion time in vivo: is the diffusion in white matter restricted?" Magnetic resonance imaging **27**(2): 176-187.

Nilsson, M., F. Szczepankiewicz, A. Ahlgren, J. d. A. Martins, S. Lasic, C.-F. Westin and D. Topgaard (2018). "An open-source framework for analysis of multidimensional diffusion MRI data implemented in MATLAB. In Proceedings of the 26th Annual Meeting of ISMRM, Paris, France, 2018. p. 5355.".

Nilsson, M., F. Szczepankiewicz, D. van Westen and O. Hansson (2015). "Extrapolation-Based References Improve Motion and Eddy-Current Correction of High B-Value DWI Data: Application in Parkinson's Disease Dementia." PloS one **10**(11): e0141825.

Nilsson, M., D. van Westen, F. Ståhlberg, P. C. Sundgren and J. Lätt (2013). "The role of tissue microstructure and water exchange in biophysical modelling of diffusion in white matter." MAGMA **26**(4): 345-370.

Novikov, D. S., J. H. Jensen, J. A. Helpern and E. Fieremans (2014). "Revealing mesoscopic structural universality with diffusion." Proceedings of the National Academy of Sciences **111**(14): 5088-5093.

Novikov, D. S., S. N. Jespersen, V. G. Kiselev and E. Fieremans (2016). "Quantifying brain microstructure with diffusion MRI: Theory and parameter estimation." arXiv preprint arXiv:1612.02059.

Novikov, D. S., V. G. Kiselev and S. N. Jespersen (2018a). "On modeling." Magnetic resonance in medicine.

Novikov, D. S., J. Veraart, I. O. Jelescu and E. J. N. Fieremans (2018b). "Rotationally-invariant mapping of scalar and orientational metrics of neuronal microstructure with diffusion MRI." **174**: 518-538.

Öngür, D., A. T. Ferry and J. L. Price (2003). "Architectonic subdivision of the human orbital and medial prefrontal cortex." Journal of Comparative Neurology **460**(3): 425-449.

Pierpaoli, C., P. Jezzard, P. J. Basser, A. Barnett and G. Di Chiro (1996). "Diffusion tensor MR imaging of the human brain." Radiology **201**(3): 637-648.

Qin, W., C. Shui Yu, F. Zhang, X. Y. Du, H. Jiang, Y. Xia Yan and K. Cheng Li (2009). "Effects of echo time on diffusion quantification of brain white matter at 1.5 T and 3.0 T." Magnetic resonance in medicine **61**(4): 755-760.

Reisert, M., E. Kellner, B. Dhital, J. Hennig and V. G. J. N. Kiselev (2017). "Disentangling micro from mesostructure by diffusion MRI: A Bayesian approach." **147**: 964-975.

Ronen, I., S. Moeller, K. Ugurbil and D.-S. Kim (2006). "Analysis of the distribution of diffusion coefficients in cat brain at 9.4 T using the inverse Laplace transformation." Magnetic resonance imaging **24**(1): 61-68.

Sagi, Y., I. Tavor, S. Hofstetter, S. Tzur-Moryosef, T. Blumenfeld-Katzir and Y. Assaf (2012). "Learning in the fast lane: new insights into neuroplasticity." Neuron **73**(6): 1195-1203.

Scholz, J., M. C. Klein, T. E. Behrens and H. Johansen-Berg (2009). "Training induces changes in white-matter architecture." Nature neuroscience **12**(11): 1370.

Sjölund, J., F. Szczepankiewicz, M. Nilsson, D. Topgaard, C.-F. Westin and H. Knutsson (2015). "Constrained optimization of gradient waveforms for generalized diffusion encoding." Journal of Magnetic Resonance **261**: 157-168.

Sotiropoulos, S. N., T. E. Behrens and S. Jbabdi (2012). "Ball and rackets: inferring fiber fanning from diffusion-weighted MRI." Neuroimage **60**(2): 1412-1425.

Stikov, N., J. S. Campbell, T. Stroh, M. Lavelée, S. Frey, J. Novek, S. Nuara, M.-K. Ho, B. J. Bedell and R. F. Dougherty (2015). "In vivo histology of the myelin g-ratio with magnetic resonance imaging." Neuroimage **118**: 397-405.

Surova, Y., F. Szczepankiewicz, J. Lätt, M. Nilsson, B. Eriksson, A. Leemans, O. Hansson, D. van Westen and C. Nilsson (2013). "Assessment of global and regional diffusion changes along white matter tracts in parkinsonian disorders by MR tractography." PloS one **8**(6): e66022.

Swieten, J. v., J. V. D. HOUT, B. V. KETEL, A. Hijdra, J. Wokke and J. v. Gijn (1991). "Periventricular lesions in the white matter on magnetic resonance imaging in the elderly: a morphometric correlation with arteriolosclerosis and dilated perivascular spaces." Brain **114**(2): 761-774.

Szafer, A., J. Zhong, A. W. Anderson and J. C. Gore (1995). "Diffusion-weighted imaging in tissues: Theoretical models." NMR in Biomedicine **8**(7): 289-296.

Szczepankiewicz, F., S. Lasič, D. van Westen, P. C. Sundgren, E. Englund, C.-F. Westin, F. Ståhlberg, J. Lätt, D. Topgaard and M. Nilsson (2015). "Quantification of microscopic diffusion anisotropy disentangles effects of orientation dispersion from microstructure: Applications in healthy volunteers and in brain tumors." Neuroimage **104**: 241-252.

Szczepankiewicz, F. and M. Nilsson (2018). "Maxwell-compensated waveform design for asymmetric diffusion encoding. Accepted for the 24th Annual Meeting of ISMRM, Paris, France, 2018.
."

Szczepankiewicz, F., J. Sjölund, F. Ståhlberg, J. Lätt and M. Nilsson (2016a). "Whole-brain diffusional variance decomposition (DIVIDE): Demonstration of technical feasibility at clinical MRI systems." arXiv preprint arXiv:1612.06741.

Szczepankiewicz, F., D. van Westen, E. Englund, C.-F. Westin, F. Ståhlberg, J. Lätt, P. C. Sundgren and M. Nilsson (2016b). "The link between diffusion MRI and tumor heterogeneity: Mapping cell eccentricity and density by diffusional variance decomposition (DIVIDE)." NeuroImage **142**: 522-532.

Szczepankiewicz, F., C.-F. Westin, F. Ståhlberg, J. Lätt and M. Nilsson (2016c). "Minimum Number of Diffusion Encoding Directions Required to Yield a Rotationally Invariant Powder Average Signal in Single and Double Diffusion Encoding. In Proceedings of the 24th Annual Meeting of ISMRM, Singapore, 2016. p. 2065.".

Tariq, M., T. Schneider, D. C. Alexander, C. A. G. Wheeler-Kingshott and H. Zhang (2016). "Bingham–NODDI: mapping anisotropic orientation dispersion of neurites using diffusion MRI." NeuroImage **133**: 207-223.

Tax, C. M., U. S. Rudrapatna, T. Witzel and D. K. Jones (2017). "Disentangling in two dimensions in the living human brain: Feasibility of relaxometry-diffusometry using ultra-strong gradients. In Proceedings of the 25th Annual Meeting of ISMRM, Honolulu, Hawaii, 2017. p. 838.".

Topgaard, D. (2016). NMR methods for studying microscopic diffusion anisotropy. Diffusion NMR in Confined Systems: Fluid Transport in Porous Solids and Heterogeneous Materials, Royal Society of Chemistry Cambridge, UK.

Topgaard, D. (2017). "Multidimensional diffusion MRI." Journal of Magnetic Resonance **275**: 98-113.

Van Cauter, S., J. Veraart, J. Sijbers, R. R. Peeters, U. Himmelreich, F. De Keyzer, S. W. Van Gool, F. Van Calenbergh, S. De Vleeschouwer and W. Van Hecke (2012). "Gliomas: diffusion kurtosis MR imaging in grading." Radiology **263**(2): 492-501.

Veraart, J., E. Fieremans and D. Novikov (2017a). "Universal Power-Law Scaling of Water Diffusion in Human Brain Defines What We See with Diffusion MRI. In Proceedings of the 25th Annual Meeting of ISMRM, Honolulu, HI, USA, 2017. p. 282.".

Veraart, J., D. S. Novikov and E. Fieremans (2017b). "TE dependent Diffusion Imaging (TEdDI) distinguishes between compartmental T2 relaxation times." Neuroimage.

Wansapura, J. P., S. K. Holland, R. S. Dunn and W. S. Ball Jr (1999). "NMR relaxation times in the human brain at 3.0 tesla." Journal of Magnetic Resonance Imaging: An Official Journal of the International Society for Magnetic Resonance in Medicine **9**(4): 531-538.

Weigel, M. and J. Hennig (2006). "Contrast behavior and relaxation effects of conventional and hyperecho-turbo spin echo sequences at 1.5 and 3 T." Magnetic resonance in medicine **55**(4): 826-835.

Wen, Q., D. A. Kelley, S. Banerjee, J. M. Lupo, S. M. Chang, D. Xu, C. P. Hess and S. J. Nelson (2015). "Clinically feasible NODDI characterization of glioma using multiband EPI at 7 T." NeuroImage: Clinical **9**: 291-299.

Werring, D., D. Brassat, A. Droogan, C. Clark, M. Symms, G. Barker, D. MacManus, A. Thompson and D. Miller (2000). "The pathogenesis of lesions and normal-appearing white matter changes in multiple sclerosis: a serial diffusion MRI study." Brain **123**(8): 1667-1676.

Westin, C.-F., H. Knutsson, O. Pasternak, F. Szczepankiewicz, E. Özarslan, D. van Westen, C. Mattisson, M. Bogren, L. J. O'Donnell and M. Kubicki (2016). "Q-space trajectory imaging for multidimensional diffusion MRI of the human brain." NeuroImage **135**: 345-362.

White, N. S., T. B. Leergaard, H. D'Arceuil, J. G. Bjaalie and A. M. Dale (2013). "Probing tissue microstructure with restriction spectrum imaging: histological and theoretical validation." Human brain mapping **34**(2): 327-346.

Whittall, K. P., A. L. Mackay, D. A. Graeb, R. A. Nugent, D. K. Li and D. W. Paty (1997). "In vivo measurement of T2 distributions and water contents in normal human brain." Magnetic Resonance in Medicine **37**(1): 34-43.

Xu, J., H. Li, K. D. Harkins, X. Jiang, J. Xie, H. Kang, M. D. Does and J. C. Gore (2014). "Mapping mean axon diameter and axonal volume fraction by MRI using temporal diffusion spectroscopy." NeuroImage **103**: 10-19.

Yablonskiy, D. A., G. L. Bretthorst and J. J. Ackerman (2003). "Statistical model for diffusion attenuated MR signal." Magnetic resonance in medicine **50**(4): 664-669.

Zhang, H., P. L. Hubbard, G. J. Parker and D. C. J. N. Alexander (2011). "Axon diameter mapping in the presence of orientation dispersion with diffusion MRI." **56**(3): 1301-1315.

Zhang, H., T. Schneider, C. A. Wheeler-Kingshott and D. C. Alexander (2012). "NODDI: practical in vivo neurite orientation dispersion and density imaging of the human brain." Neuroimage **61**(4): 1000-1016.

# Supporting information

## 1) Using a 'ball' or a 'tortuosity' constraint for the shape of the 'non-stick' compartment only had a small impact on estimated 'stick' fraction values

The models used in this work (manuscript Table 1) enforced the 'ball' constraint ($D_\Delta = 0$) of isotropic diffusion in the 'non-stick' compartment, motivated by an expected high degree of orientation dispersion of anisotropic structures in the target regions (manuscript figure 1). The constraint sets C3 and C5 differ in this manner from the sets used by SMT and NODDI, which employ a 'tortuosity' constraint for the shape of the 'non-stick' compartment (Zhang et al. 2012, Kaden et al. 2016), according to

$$D_\Delta = f_S / (3 - 2 \cdot f_S). \qquad \text{Eq. 1}$$

Here, we sought to demonstrate that the 'ball' constraint was a good approximation to the 'tortuosity' constraint in the investigated regions. For this purpose, we defined two sets of constraints: 'C3 ball', which is simply C3 with the 'ball' constraint as used in the manuscript, and 'C3 tortuosity', which is C3 together with Eq. 1. The two corresponding models were fit, for all subjects, to the combined b-tensor and multi-echo data in the group A ROIs.

Using either the 'ball' or 'tortuosity' constraint yielded similar values of the 'stick' fraction (Fig. 1). 'C3 ball' generally yielded somewhat higher 'stick' fraction values than 'C3 tortuosity', which was expected since the latter set allows some diffusion anisotropy to be captured also in the 'non-stick' compartment. The differences were overall small, however, and smaller in regions expected to feature a higher degree of orientation dispersion: from approximately 1% in the insular cortex to approximately 6% in the anterior corona radiata.

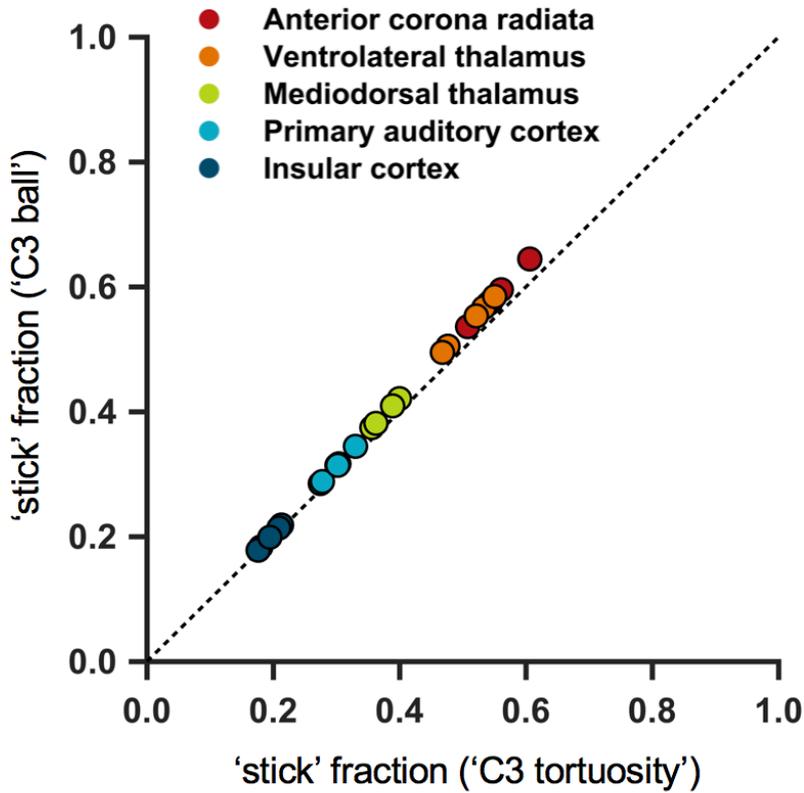

**Figure 1 – Using either a 'ball' or a 'tortuosity' constraint for the 'non-stick' compartment yielded similar 'stick' fraction values in the group A ROIs.** Using constraint set C3 with a 'ball' constraint ($D_\Delta = 0$) yielded only slightly higher 'stick' fraction values than using a 'tortuosity' constraint (Eq. 1). The differences were smallest in the insular cortex ($\approx$ 1%) and largest in the anterior corona radiata ($\approx$ 6%).

**2) The goodness of fit-plots represented the ranges of data-compatible 'stick' fractions in the cortex well, despite a free CSF fraction**

The minimally constrained model featured a free CSF fraction in the cortical regions of group A, which should reduce the model's overall precision and might cause the visualization of data-compatible 'stick' fractions (manuscript Fig. 4A, second row) to miss local minima. To verify that the plots yielded good representations of the ranges of data-compatible 'stick' fraction, we repeated the procedure described in Method section 3.4 for 16 fixed values of the CSF fraction between zero and 0.1.

Figure 2 (upper row) shows, for both cortical regions and one example subject, two-dimensional equivalents of the normalized residual variance (manuscript Eq. 18) plots for the non-cortical regions in manuscript figure 4 (upper row). Comparing these to the results obtained using a free CSF fraction for the same subject (Fig. 2, second row) indicates that the latter curves yield a good representation of the range of data-compatible 'stick' fractions. In particular, the reduced precision from using a free CSF fraction did not obscure local minima of acceptable 'stick' fractions.

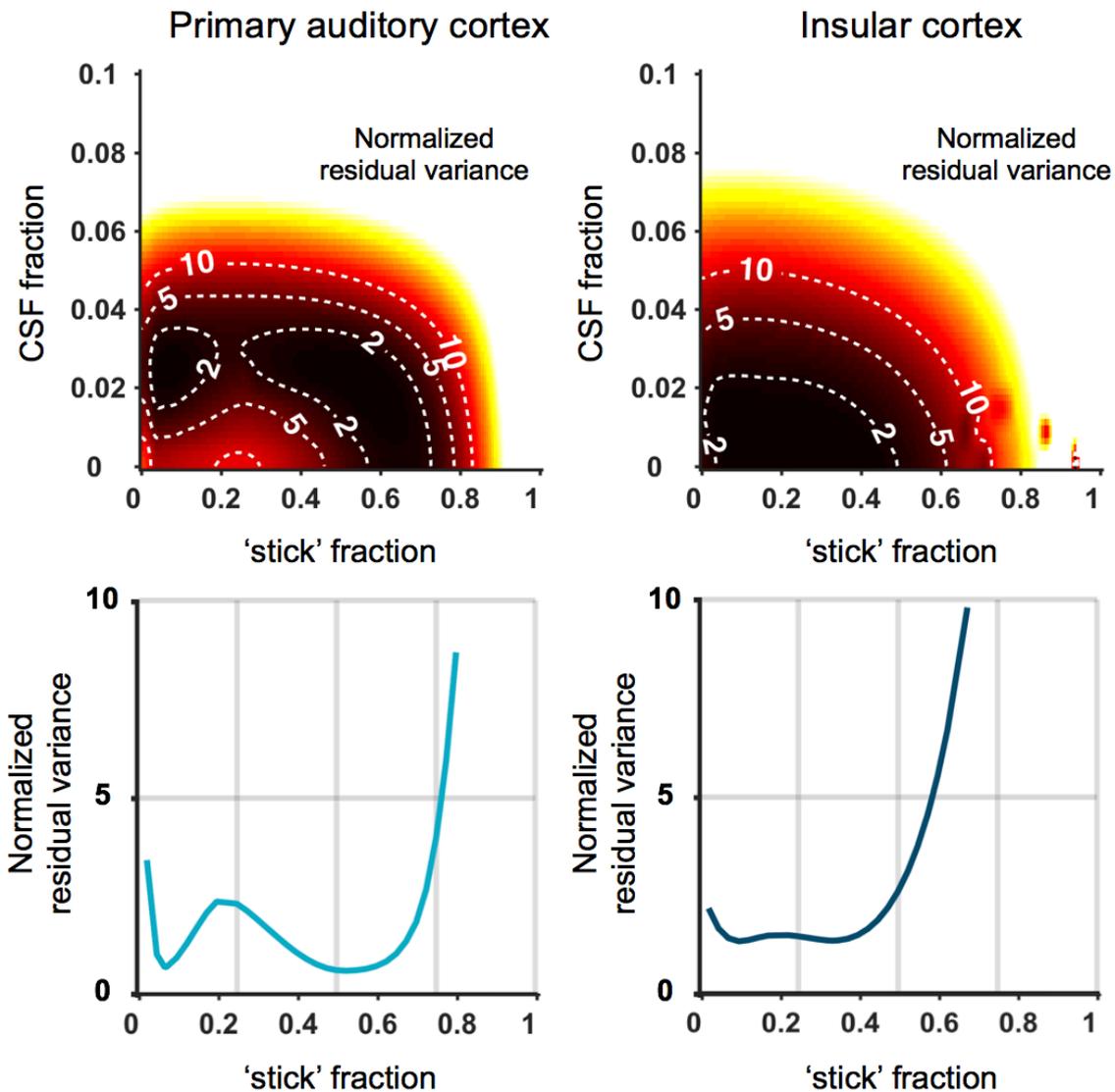

**Figure 2 – The goodness of fit-plots yielded representative ranges of data-compatible 'stick' fractions in the cortex despite the free CSF fraction.** The figure compares plots of the normalized residual variance (manuscript Eq. 18) obtained either with iterative fixing of both the 'stick' and CSF fractions (upper row), or with fixing of the 'stick' fraction only (lower row). We conclude that the latter procedure, which was used for manuscript figure 4, yielded representative ranges of data-compatible 'stick' fractions.

**3) Using an LTE waveform that was either 'tuned' or 'detuned' with respect to the STE waveform had no systematic impact on the signal**

The gradient waveforms used in the study for LTE were 'detuned' with respect to those used for STE in that they featured a different frequency ('spectral') content. It has been demonstrated ex vivo that such a discrepancy may confound time-dependent diffusion for microscopic anisotropy (Lundell et al. 2017). To investigate the impact of this effect, we acquired additional LTE data in group A using a waveform that was 'tuned' to the same spectral content as the waveform used for STE. The acquisition employed the LTE protocol in manuscript Table 1 for $b \leq 0.5$ ms/μm$^2$.

The 'tuned' and 'detuned' LTE gradient waveforms yielded similar signal values. Figure 3 compares the (powder-averaged) signals at $b = 0.1$ ms/μm$^2$ and $b = 0.5$ ms/μm$^2$ for all regions and subjects. No systematic differences were observed for any region or b-value, indicating a small impact from their respective differences in spectral content. Subject deviation from the line of identity ('tuned' = 'detuned', dashed line) was often, but not always, within one signal standard deviation due to noise (error bars). From this, we conclude that the differences in spectral content between LTE and STE had a minimal impact on our data.

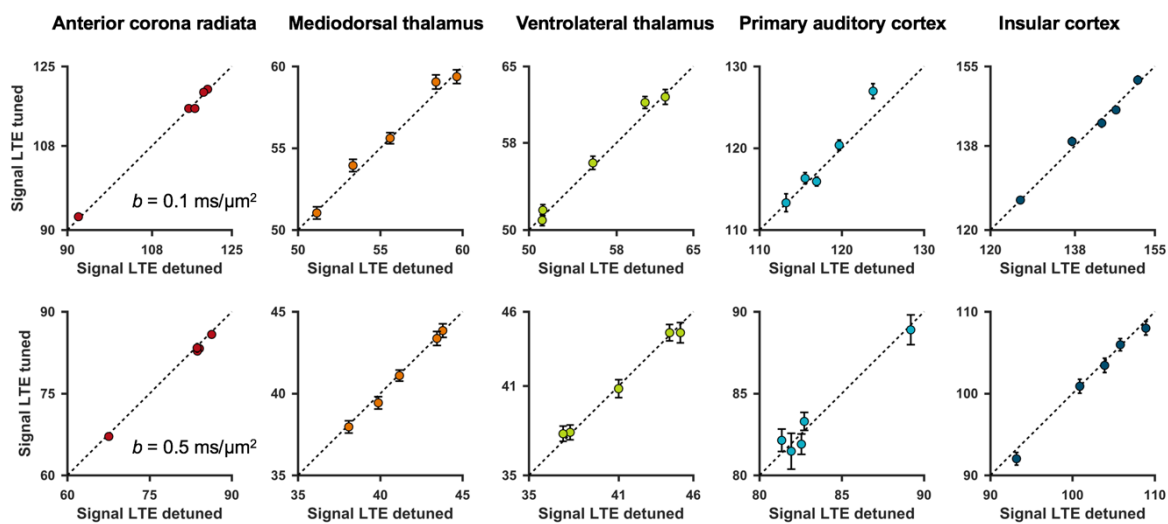

**Figure 3 – Using an LTE waveform that was either 'tuned' or 'detuned' with respect to the STE waveform yielded similar signals.** The figure compares powder-averaged 'tuned' and 'detuned' LTE signals for two b-values (rows) in all regions (columns) and subjects (points). Dashed lines indicate lines of identity and error bars denote signal standard deviations due to noise. No systematic difference was observed and the deviations were generally small.

**4) Extending the multi-echo LTE acquisition to high b-values could yield precision to separate compartment T2 times in the healthy brain.**

Although compartment T2 times may be more easily separated if multi-echo data are acquired for high b-values (Veraart et al. 2017b), the current protocol was limited to $b \leq 0.5$ ms/$\mu$m$^2$ for sufficient SNR in deep gray matter using 2 mm isotropic voxels. To explore the impact of higher b-values on the precision of the minimally constrained model, we simulated extended datasets based on three sets of ground truth parameters from each region (manuscript Fig. 4A, one subject). The extended protocol equaled that of group A (manuscript Table 2) but with an additional 192 LTE samples, using TE = [69 90 120 150] ms for b-values of 1, 1.5 and 2 ms/$\mu$m$^2$ averaged across 12, 16 and 20 directions, respectively. Rician noise was added using estimates of signal variance due to noise, but rescaled to reflect 2.5 mm isotropic voxels, resulting in a minimal lowest SNR of approximately 2 in the mediodorsal thalamus. The precision was visually assessed using normalized residual variances (manuscript Eq. 18) and the procedure described in Method section 3.4, fixing $f_C$ to zero.

    The result in Fig. 4 suggests that the extended protocol (solid lines) would yield precision to the 'stick' fraction of the minimally constrained model for most regions (rows) and ground truth solutions (columns; vertical dashed lines). In the primary auditory cortex, however, estimates were either unprecise or biased, likely due to partial volume effects with CSF (ground truth $f_C$ was clearly non-zero only in this region). The current protocol (dashed lines) yielded a low precision similar to for the actual data (manuscript Fig. 4A).

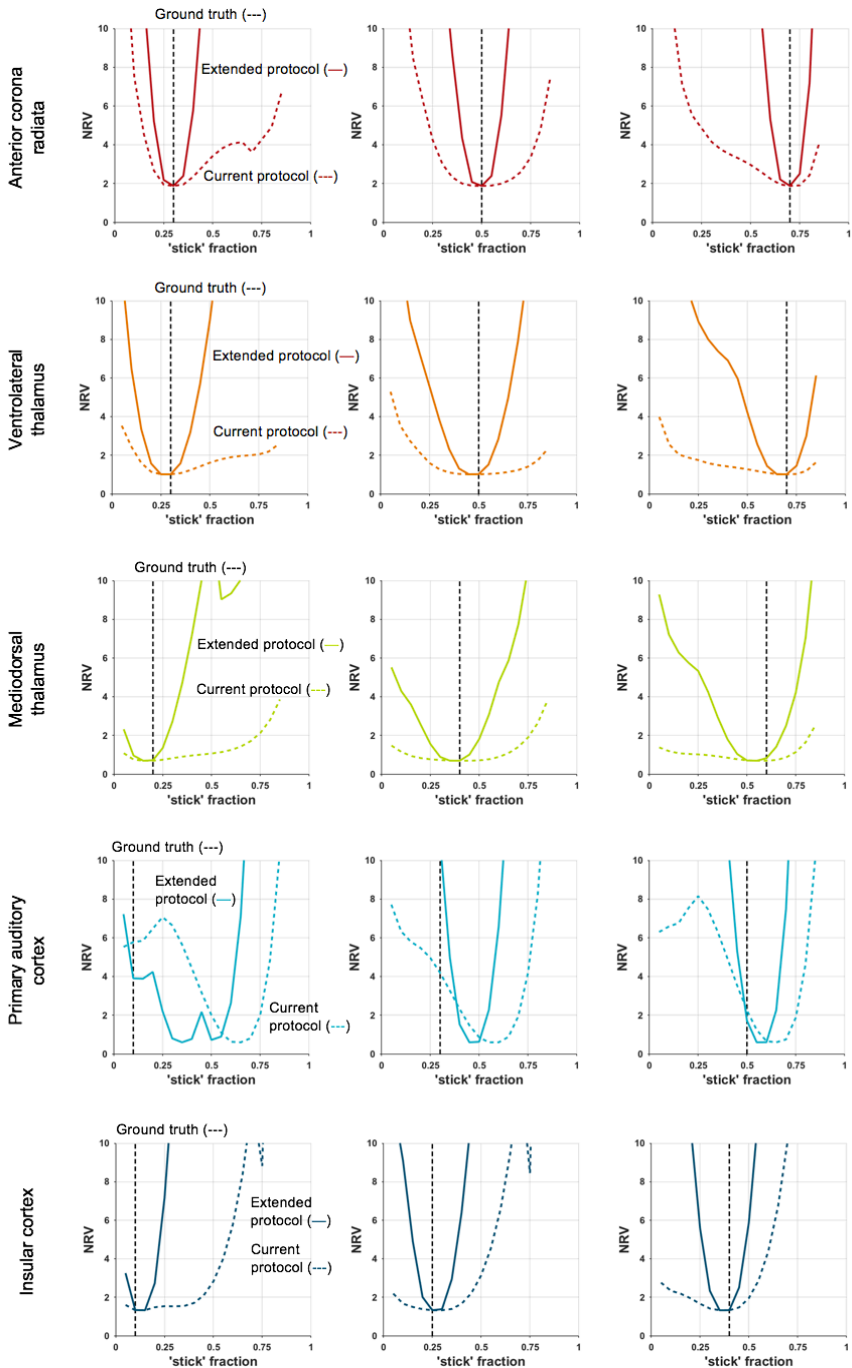

**Figure 4 – Extending the protocol to multi-echo data for high b-values increased the precision of the minimally constrained model.** In most regions (rows) and ground truth solutions (columns; vertical dashed lines), the extended protocol (solid lines) yielded a higher precision to estimate the 'stick' fraction and compartment-specific T2 times compared to the current protocol (dashed lines). The issues with accuracy and precision in the primary auditory cortex could reflect CSF. Ground truth 'stick' fractions were, for the five regions in order, [0.3 0.5 0.7], [0.3 0.5 0.7], [0.2 0.4 0.6], [0.1 0.3 0.5] and [0.1 0.25 0.4].

# References

Kaden, E., N. D. Kelm, R. P. Carson, M. D. Does and D. C. Alexander (2016). "Multi-compartment microscopic diffusion imaging." <u>NeuroImage</u> **139**: 346-359.

Lundell, H., M. Nilsson, T. B. Dyrby, G. J. Parker, P. L. H. Cristinacce, F. Zhou, D. Topgaard and S. Lasic (2017). "Microscopic anisotropy with spectrally modulated q-space trajectory encoding. In proceedings for the 25th Annual Meeting of ISMRM, Honolulu, Hawaii, USA, 2017. p. 1086.".

Veraart, J., D. S. Novikov and E. Fieremans (2017). "TE dependent Diffusion Imaging (TEdDI) distinguishes between compartmental T2 relaxation times." <u>Neuroimage</u>.

Zhang, H., T. Schneider, C. A. Wheeler-Kingshott and D. C. Alexander (2012). "NODDI: practical in vivo neurite orientation dispersion and density imaging of the human brain." <u>Neuroimage</u> **61**(4): 1000-1016.